\documentclass[review]{elsarticle}

\usepackage{amssymb}
\usepackage{amsmath,amssymb,amsfonts}
\usepackage{algorithmic}
\usepackage{graphicx}
\usepackage{textcomp}
\usepackage{xcolor}
\usepackage{threeparttable}
\usepackage{multirow}
\usepackage{subfigure}
\usepackage{algorithm}
\usepackage{mathrsfs}
\usepackage{adjustbox}

\usepackage{array}
\newcommand{\PreserveBackslash}[1]{\let\temp=\\#1\let\\=\temp}
\newcolumntype{C}[1]{>{\PreserveBackslash\centering}p{#1}}
\newcolumntype{R}[1]{>{\PreserveBackslash\raggedleft}p{#1}}
\newcolumntype{L}[1]{>{\PreserveBackslash\raggedright}p{#1}}

\journal{Journal of \LaTeX\ Templates}

\biboptions{authoryear}

\begin{document}

\begin{frontmatter}

\title{Cascaded Context Enhancement Network for Automatic Skin Lesion Segmentation}

\author[1]{Ruxin Wang}
\ead{rx.wang@siat.ac.cn}
\author[1,2]{Shuyuan Chen}
\ead{sa517020@mail.ustc.edu.cn}
\author[1]{Chaojie Ji}
\ead{cj.ji@siat.ac.cn}
\author[1]{Ye Li\corref{mycorrespondingauthor}}
\cortext[mycorrespondingauthor]{Corresponding author}
\ead{ye.li@siat.ac.cn}
\address[1]{Shenzhen Institutes of Advanced Technology, Chinese Academy of Sciences, China}
\address[2]{School of Software Engineering, University of Science and Technology of China, China}

\begin{abstract}

Skin lesion segmentation is an important step for automatic melanoma diagnosis. Due to the non-negligible diversity of lesions from different patients, extracting powerful context for fine-grained semantic segmentation is still challenging today. Although the deep convolutional neural network (CNNs) have made significant improvements on skin lesion segmentation, they often fail to reserve the spatial details and long-range dependencies context due to consecutive convolution striding and pooling operations inside CNNs. In this paper, we formulate a cascaded context enhancement neural network for automatic skin lesion segmentation. A new cascaded context aggregation (CCA) module with a gate-based information integration approach is proposed to sequentially and selectively aggregate original image and multi-level features from the encoder sub-network. The generated context is further utilized to guide discriminative features extraction by the designed context-guided local affinity (CGL) module. Furthermore, an auxiliary loss is added to the CCA module for refining the prediction. In our work, we evaluate our approach on four public skin dermoscopy image datasets. The proposed method achieves the Jaccard Index (JA) of $87.1\%$, $80.3\%$, $83.4\%$ and $86.6\%$ on ISIC-2016, ISIC-2017, ISIC-2018, and PH2 datasets, which are higher than other state-of-the-art models respectively.

\end{abstract}

\begin{keyword}

Convolutional neural network \sep Deep learning \sep Dermoscopy image \sep Skin lesion segmentation

\end{keyword}

\end{frontmatter}


\section{Introduction}

Skin cancer is one of the most common human malignancies globally accounting for at least $40\%$ of cases \citep{barata2017development,cakir2012epidemiology}. In the United States, there are more than five million new cases each year \citep{siegel2016cancer}. It has been a major public health problem and the incidence of skin cancer has been gradually rising \citep{guy2015prevalence}. Among them, melanoma is a highly malignant tumor, which is prone to metastasis and endangers the lives of patients. Meanwhile, melanoma is a kind of cancer with a high survival rate if treated promptly by chemotherapy, radiation therapy, and surgery at its early stage \citep{leachman2016methods}. Therefore, early accurate diagnosis and screening are the keys to enhance the cure rate. As pigmented lesions occurring on the surface of the skin, melanoma is amenable to early detection by expert visual inspection. Dermoscope imaging technique is one of the effective non-invasive skin cancer detection methods, which can enhance the visualization of deeper levels of the skin and assist clinicians in diagnosis \citep{kittler2002diagnostic,silveira2009comparison}. Nevertheless, melanoma and other pigmented skin lesions are similar in color and texture. And they have obvious individual differences. Therefore, only experienced dermatologists can accurately identify them, while clinicians with limited experience may affect the optimal timing of treatment. Also, the diagnostic process is laborious and time-consuming.

\begin{figure*}[h]
	\centerline{\includegraphics[width=0.8\textwidth]{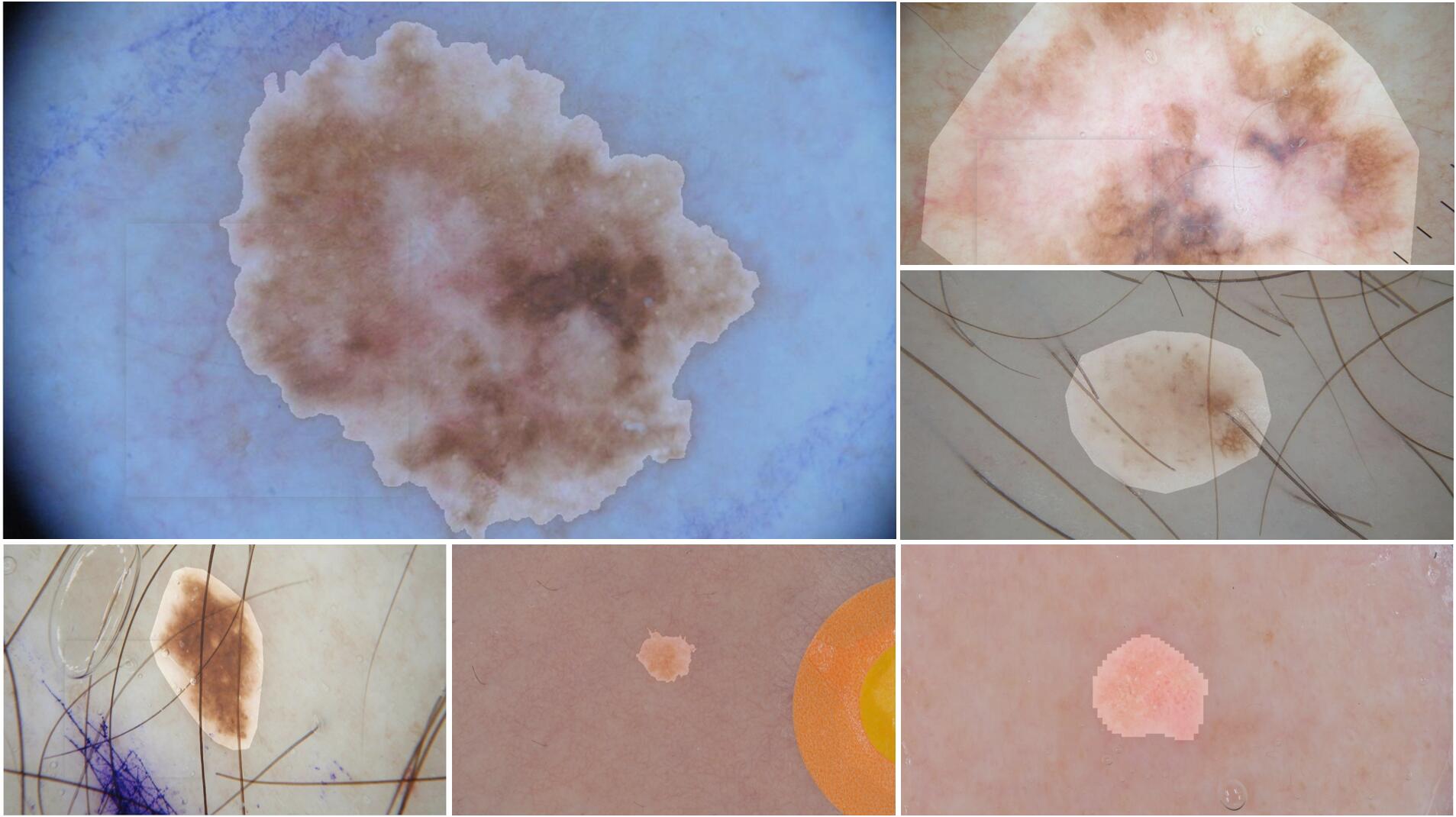}}
	\caption{Some examples for challenges (\emph{e.g.} low contrast, irregular shape, fuzzy borders, and some artifacts, etc.) of automated melanoma recognition. The highlighted area in each image indicates the lesion part.}
	\label{fig:1}
\end{figure*}

With the development of Artificial Intelligence, the computer-assisted diagnosis system can effectively avoid many of the above problems and help doctors improve the accuracy and efficiency of diagnosis. For automatic melanomas recognition, however, lesion segmentation as a fundamental component remains a challenging task \citep{abbas2013pattern,yuan2017automatic}. First, most melanomas consist of various colors from shades of brown to black, the size and shape of the lesions are also significantly distinct. Second, the contrast between the boundary part of the lesion and the normal skin is low and the difference is not obvious, which increases the difficulty of segmentation. At last, some artifacts including hair, ruler marks, and color calibration also affect the accuracy of recognition. Figure \ref{fig:1} shows some actual dermoscopy image examples with various lesion areas.

Over the past decade, large numbers of automatic analysis algorithms for skin lesion segmentation have been proposed \citep{ganster2001automated,chung2000segmenting,xie2013automatic,tang2009multi}, which can be roughly classified into three categories: thresholding-based \citep{ganster2001automated}, edge-based \citep{chung2000segmenting}, and region-based methods \citep{yuan2009narrow}. Although these early attempts improve the accuracy of automatic lesion recognition through extracting pixel or region features, they still have some common defects: (1) low-level hand-crafted features (\emph{e.g.} color, texture) restrict the performance of prediction in that other available cues in the original image are neglected. (2) The robustness to artifacts, image quality, and contrast is low, which heavily depends on effective pre-processing.

Recently, compared with the approaches based on the hand-crafted features, significantly improved performance has been achieved on skin lesion segmentation by deep convolutional neural network (CNN) \citep{esteva2017dermatologist,yuan2017improving,bi2019step}, which mainly adopts the encoder-decoder structure, \emph{e.g.} U-Net \citep{ronneberger2015u}, fully convolutional network (FCN) \citep{long2015fully}. In the encoder stage, the high-level semantic feature representations are obtained through multiple convolution operations, then the top features are embedded into the decoder sub-network and generate the predicted segmentation mask. However, although high-level features have more abundant semantic features, much spatial detail is lost caused by consecutive convolution striding and pooling operations, which yields insufficient prediction performances. Additionally, capturing the long-range dependencies context plays an important role in promoting the performance of the segmentation. Many recent well-known works present different approaches to overcome these problems and get the informative context, such as dilated convolution \citep{lessmann2017automatic}, some non-local methods \citep{luo2017non}, and skip connection \citep{ronneberger2015u}. In order to obtain more semantic information, these approaches mainly focus on enlarging the local respective field in each stage of the encoder part, or bridging the feature maps from the encoder part with the same spatial resolution during the decoding process. However, for complex and fine lesions, they fail to reserve sufficient spatial details of the target and long-range dependencies contextual information. Overall, how to learn richer context and is still a challenge of the segmentation algorithm.

To address the above problems, we develop a novel neural network framework for skin lesion segmentation following the encoder-decoder structure. To embed more spatial details into high-level semantic features and encode long-range dependencies context, a cascaded context aggregation module (CCA) is designed by gradually and selectively aggregating original image and features from the encoder sub-network. The generated context of the CCA module is employed as a global context to guide the fine-grained mining of local context relationships. In order to effectively integrate more global and complementary information for segmentation prediction, a new context-guided local affinity (CGL) module is proposed. Furthermore, an auxiliary loss module is added to the generated global context to refine the prediction performances. Finally, we evaluate our model on four public datasets, \emph{i.e.} ISIC-2016 \citep{gutman2016skin}, ISIC-2017 \citep{codella2018skin}, ISIC-2018 \citep{codella2019skin}, and PH2 \citep{mendoncca2013ph}. The results demonstrate that our work has significant performance for skin lesion segmentation than other state-of-the-art methods. Our main contributions of this paper can be summarized as follows:

(1) We present a cascaded context enhancement neural network, which employs a cascaded context aggregation module and a context-guided local affinity module to produce richer contextual information for skin lesion segmentation.

(2) We design a cascaded context aggregation module to embed additional spatial structural details into high-level features and generate richer non-local context. By designing a new gate-based information integration approach, multi-level features are well integrated.

(3) We propose a context-guided local affinity module that is used to extract discriminative features by leveraging the generated context to guide local affinity computation.

(4) We achieve outstanding state-of-the-art skin lesion segmentation performance on four public dermoscopy image datasets. The experimental results convince the efficiency of the proposed method.

The remainder of this paper is organized as follows: Section 2 is the related work with respect to skin lesion segmentation. In Section 3, we describe the details of our proposed segmentation neural network. Section 4 presents experimental results with different methods on four public datasets. And related analysis of our method is discussed. Finally, Section 5 concludes this paper and prospects some future works.

\section{Related work}

\subsection{Encoder-decoder segmentation framework}

In recent years, deep learning algorithms have wide applications in computer vision. Semantic segmentation is one of the fundamental vision tasks in daily life, which assigns semantic labels to every pixel in an image. Deep learning methods \citep{chen2017deeplab,lin2018multi} based on the FCN \citep{long2015fully} show striking improvement using encoder-decoder framework for many segmentation tasks. In the encoder sub-network, image content is encoded by multiple convolutional layers from low-level to high-level. And in the decoder part, the prediction mask is obtained by multiple deconvolutional (upsampling/uppooling) layers. In particular, image feature coding and context extraction are the keys for segmentation tasks. Some outstanding works have been developed by leveraging multi-scale convolution and attention mechanisms. For example, DeepLab \citep{chen2017deeplab,chen2017rethinking} designed an atrous spatial pyramid pooling module to encode the multi-scale contextual information by parallel dilated convolution. PSPNet \citep{zhao2017pyramid} adopted a pyramid pooling module to capture the rich context in the encoder part using different pyramid scales. \cite{yang2018denseaspp} proposed densely connected atrous spatial pyramid pooling by connecting each set of atrous convolutional layers in a dense way. PSANet \citep{zhao2018psanet} used the point-wise spatial attention network to learn a pixel-wise global attention map. Through effective context feature extraction, the segmentation performance of the network can be greatly improved.

\subsection{Skin lesion segmentation}

For skin lesion segmentation, many achievements have been made based on deep learning over these few years. Most of these studies follow the deep fully convolutional neural network and encoder-decoder structure for lesions segmentation. \cite{bi2017dermoscopic} leveraged a multi-stage fully convolutional network to automatically segment the skin lesions, which learns complementary visual features of lesions. \cite{yuan2017automatic} proposed a 19-layer deep convolutional neural network without prior knowledge of the data. And a novel loss function based on Jaccard distance is designed for sample re-weighting. \cite{xue2018adversarial} proposed a SegAN network for skin lesion segmentation based on the adversarial neural network. An FCN in the last layer is employed to generate segmentation label maps, and an adversarial critic network with a multi-scale L1 loss function to capture long- and short-range spatial relationships between pixels. \cite{chen2018multi} adopted a multi-task learning framework to process the skin lesion segmentation and classification simultaneously, which employs the features from different tasks and obtains richer contextual information. \cite{li2018dense} designed a new dense deconvolutional network for skin lesion segmentation based on residual learning. It captures fine-grained multi-scale features of the image for segmentation task by dense deconvolutional layers, chained residual pooling and auxiliary supervision. \cite{sarker2018slsdeep} presented SLSDeep network based on encoder-decoder structure. And a loss function by combining both negative log-likelihood and end point error is adopted to accurately segment lesions with sharp boundaries. However, due to constrained local receptive fields and insufficient
short-range contextual information, the performances of many
segmentation approaches are limited. Overall, effective extraction of image context is crucial to enhance lesion segmentation performance.

\begin{figure*}[!t]
	\centerline{\includegraphics[width=1\textwidth]{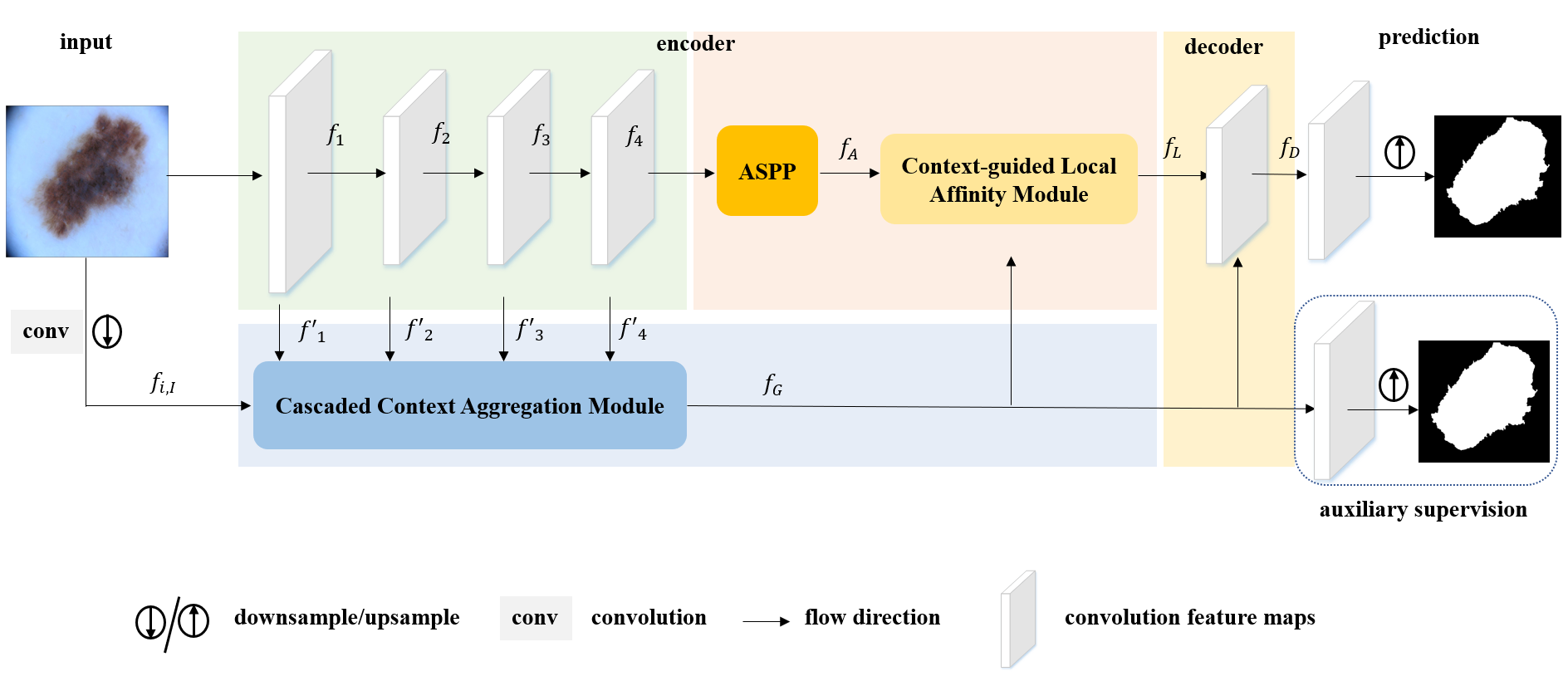}}
	\caption{Overview of the proposed model architecture. CCA module produces richer context to guide the decoder processing. CGL module is employed to further extract discriminative features by leveraging the local affinity relationship. Auxiliary supervised module uses the generated context features of CCA to refine the prediction result and assist network optimization.}
	\label{fig:2}
\end{figure*}

\section{Methodology}

In this section, we first present a general framework of our method and then the details of the CCA module and CGL module are introduced which extract powerful global context and aggregate discriminative features for fine-grained semantic segmentation. Finally, we describe network optimization.

\subsection{Overview}

The overall framework design of the proposed model is illustrated in Figure \ref{fig:2}. For simplicity, we denote training dataset $D=\{(X^{(k)},Y^{(k)}|k = 1,2,...,K\}$ of $K$ dermoscopy images, where $X^{(k)}$ indicates one annotated image with pixel-level labels $Y^{(k)}$. The label $Y_j^{(k)}$ of each pixel $j$ belongs to $\{0,1\}$, which $0$ indicates the background pixel and $1$ denotes the foreground (lesion) pixel. The goal of deep learning based on skin lesion segmentation is to learn powerful contextual information for obtaining a fine-grain prediction mask. The proposed method is based on an encoder-decoder architecture. Specifically, we employ the ResNet \citep{he2016deep} (pre-trained on ImageNet \citep{russakovsky2015imagenet}) as the backbone. And the atrous spatial pyramid pooling (ASPP) module \citep{chen2017deeplab} is adopted after the last ResNet block to generate multi-level feature representations. The ASPP module consists of four parallel atrous convolutions with different atrous rates and one global average pooling, which is effective to capture multi-scale representations and encode context. The output features of ASPP are concatenated by upsampling and one $1\times 1$ convolution (with $256$ filters). Meantime, to maintain the large spatial resolution for the segmentation task, in our work, the last two blocks in ResNet are modified with atrous convolution (atrous rate $= 2$) and remove the pooling operation. Thus, the sizes of these feature maps in each ResNet block are $1/4$, $1/8$, $1/8$ and $1/8$ of the input image. For the sake of encoding long-range dependencies context and embedding more spatial details to high-level features, a cascaded context aggregation module (CCA) is designed to sequentially enhance context features by integrating external origin image and features from each level of the encoder sub-network. Therefore, more spatial details are embedded into high-level semantic features and the context dependencies from low- to high- level are recorded and mined. Furthermore, the generated context features of the CCA are employed to guide to extract discriminative features by the context-guided local affinity module (CGL). The CGL leverages the local affinity information and forms the finer context for the decoder part. In the decoder sub-network, we aggregate the output feature maps from the CCA module and CGL module, and further integrate them using a convolution layer for final skin lesion segmentation. Meantime, an auxiliary supervised module that utilizes the generated context features is added to the framework for assisting network optimization.

\begin{figure*}[!t]
	\centerline{\includegraphics[width=1\textwidth]{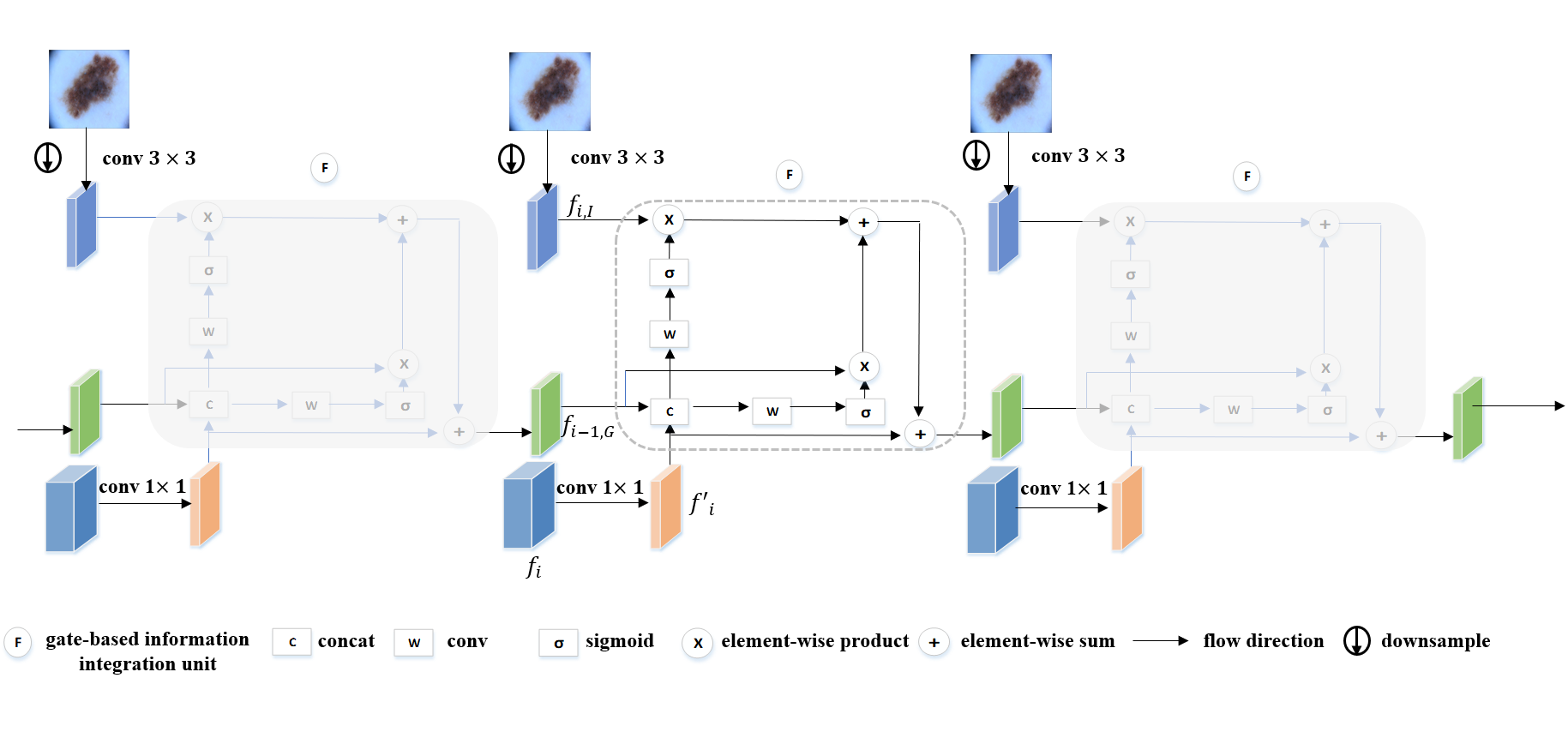}}
	\caption{The design of proposed gate-based information integration unit, where the inputs of one unit for $i$th integration are the generated context features $f_{i-1,\mathcal{G}}$ of $i-1$th integration, network features $f'_i$ of current level and the image features $f_{i,\mathcal{I}}$.}
	\label{fig:3}
\end{figure*}

\subsection{Cascaded context aggregation module}

In the encoder sub-network, the low-level features are rich in spatial details, and the high-level features have more abundant semantic features \citep{zhang2018exfuse} after consecutive pooling and convolution operations. Due to the lack of spatial detail and long-range dependencies context, however, the performance of decoding with high-level features alone is limited. In order to embed additional spatial structural details into high-level features and generate richer global context, a cascaded context aggregation (CCA) module is introduced by gradually aggregating the original image and encoder sub-network features of each stage. The CCA module combines features from low-level to high-level and external raw image information adaptively, thus enhancing the ability of the context representation.

In our CCA module, the core component is the gate-based information integration unit, which is design for sequentially and selectively aggregating the message flows from different sources. For one image $X^{(k)}$, without loss of generality, we denote the feature maps extracted from the last convolution layer of each residual convolutional block as $f_i, i\in \{1,2,3,4\}$. As illustrated in Figure \ref{fig:3}, to extract powerful high-level features and capture long-range dependencies context, we use the external image information, the feature maps of front level and current level for refining and supplementing spatial details in each stage of the encoder sub-network. A gate-based information integration unit is designed to gradually aggregate the above message flows, which contains three inputs and one output setting. Specifically, for $i$th integration, the inputs are the generated context features $f_{i-1,\mathcal{G}}$ of $i-1$th integration, network features $f'_i$ of current level and the image features $f_{i,\mathcal{I}}$. First, we obtain the output features $f'_i$ of each stage using $1\times 1$ convolution, which aims to further incorporate and reduce the dimension of features $f_i$. It can be defined as follows:
\begin{eqnarray}
&& f'_i  = \mathcal{F}(f_i; \theta_i), \quad \quad i\in\{1,2,3,4\}
\end{eqnarray}
where $\mathcal{F}$ indicates one $1\times 1$ convolution and $\theta_i$ denotes the respective parameter. And the original image $X$ is resized to the same spatial resolution as features $f'_i$ by downsampling operation. Then, features $f_{i,\mathcal{I}}$ with same channel numbers as $f'_i$ are extracted by a $3\times 3$ convolution operation:
\begin{eqnarray}
&& f_{i,\mathcal{I}}  = \mathcal{F}(Down(X^{(k)}); \theta_{i,\mathcal{I}}), \quad \quad i\in\{1,2,3,4\}
\end{eqnarray}
where $Down(\cdot)$ refers to a downsampling operator.  $\theta_{i,\mathcal{I}}$ is the convolution parameter. Thus, the addition-based fusion process in one gate-based information integration unit can be formulated as follows:
\begin{equation}
f_{i,\mathcal{G}}=\left\{
\begin{aligned}
&f'_i,  \quad\quad\quad\quad\quad\quad\quad\quad\quad\quad\quad\quad\quad\quad i=1  \\
&f'_i + a_{i,\mathcal{I}}\otimes f_{i,\mathcal{I}} + a_{i,\mathcal{G}}\otimes f_{i-1,\mathcal{G}}, \quad   i\in\{2,3,4\}
\end{aligned}
\right.
\end{equation}
where $a_{i,\mathcal{I}}$ and $a_{i,\mathcal{G}}$ denote the gate masks of features $f_{i,\mathcal{I}}$ and $f_{i-1,\mathcal{G}}$ respectively, which is used to screen important message from $f_{i,\mathcal{I}}$ and $f_{i-1,\mathcal{G}}$ adaptively. And $\otimes$ denotes the element-wise product. The gate function controls the message passing which selectively filters useful cues of front level and original image to current level. In our work, the gate function is designed using the $1\times 1$ convolution and a sigmoid activation function to produce filter masks for $f_{i,\mathcal{I}}$ and $f_{i-1,\mathcal{G}}$, which can be defined as follows:
\begin{eqnarray}
& a_{i,\mathcal{I}} = \sigma(\mathcal{F}(\mathcal{C}(f'_i, f_{i-1,\mathcal{G}});\theta_{i,\mathcal{I}}) \\
& a_{i,\mathcal{G}} = \sigma(\mathcal{F}(\mathcal{C}(f'_i, f_{i-1,\mathcal{G}});\theta_{i,\mathcal{G}})
\end{eqnarray}
where $\mathcal{C}$ refers to the concatenation operation. $\theta_{i,\mathcal{I}}$ and $\theta_{i,\mathcal{G}}$ are the parameters of the $1\times 1$ convolution. $\sigma$ indicates the element-wise sigmoid function. For the sake of simplification, we define $f_{\mathcal{G}}$ as the output of our CCA module. By the CCA module, we gradually enhance the context features and avoid introducing too much redundant content by the designed gate-based information integration unit. Meanwhile, the generated context $f_{\mathcal{G}}$ can be viewed as the global feature maps to guide the local context learning, and improve the performance of segmentation as an auxiliary supervised branch.

\begin{figure*}[htbp]
	\centerline{\includegraphics[width=1\textwidth]{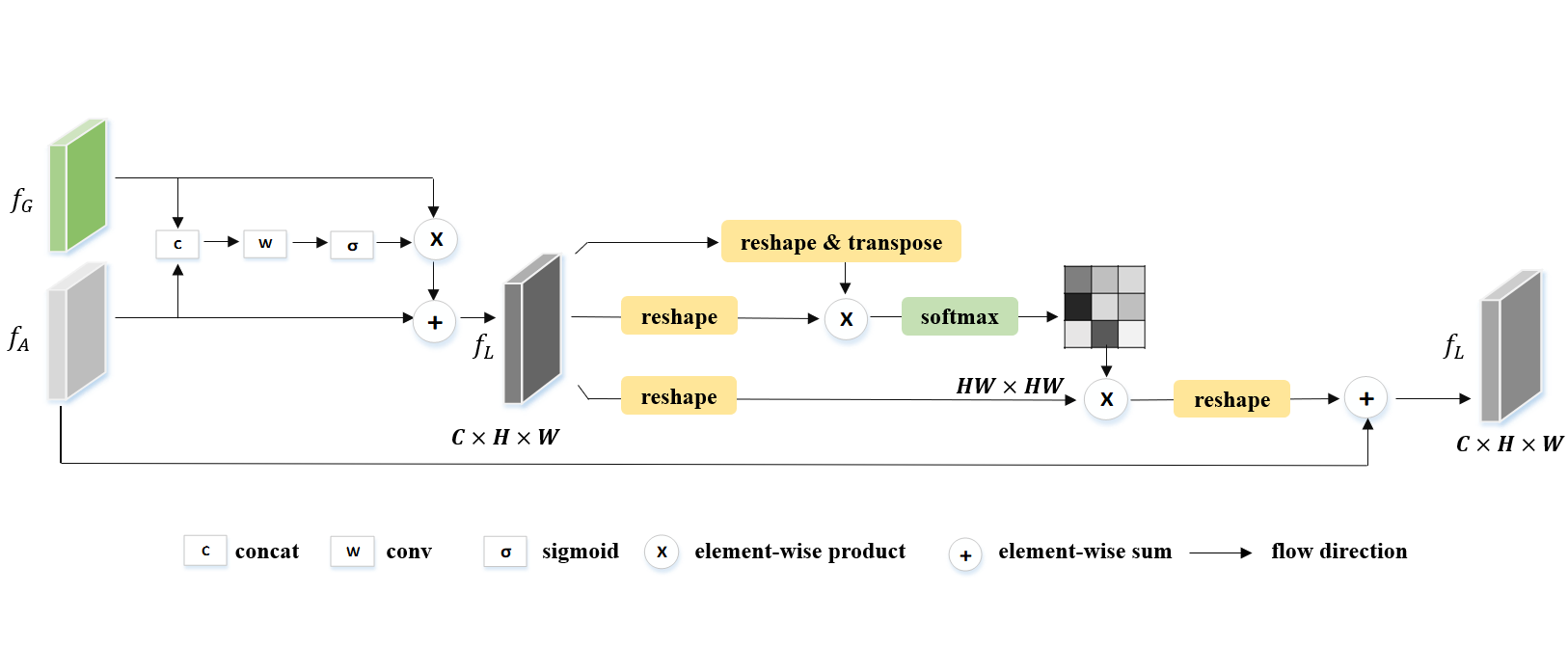}}
	\caption{The design of the proposed CGL module, where the inputs of CGL are the generated context features $f_\mathcal{G}$ and the features $f_\mathcal{A}$ of the ASPP module.}
	\label{fig:4}
\end{figure*}

\subsection{Context-guided local affinity module}

To further promote the features representation and discriminative capability, we propose a context-guided local affinity (CGL) module to capture the semantic relationship among different positions in local feature maps based on the position attention. In local feature maps, any two local positions or regions with similar features have correlated complementary improvement, which is helpful to improve the discriminative ability of the encoder features. Moreover, we employ the generated context $f_{\mathcal{G}}$ for guidance, which can effectively integrate more global and complementary cues. The structure of the CGL module is illustrated in Figure \ref{fig:4}. The CGL module enhances the ability of feature representation by encoding a wider range of semantic information into the local receptive field. First, to generate an enhanced local feature maps, we adopt the similar gate-based fusion method to aggregate the global context $f_\mathcal{G}$ and the features $f_\mathcal{A}$ of the ASPP module:
\begin{align}
& a_{\mathcal{L}} = \sigma(\mathcal{F}(\mathcal{C}(f_\mathcal{A}, f_\mathcal{G}) ;\theta_\mathcal{L})  \\
& f_\mathcal{L}= f_\mathcal{A} + a_{\mathcal{L}}\otimes f_\mathcal{G}
\end{align}
where $a_{\mathcal{L}}$ is the filter masks of generated context features $f_\mathcal{G}$. $\theta_\mathcal{L}$ denotes the parameters of $1\times 1$ convolution. Then the obtained features $f_\mathcal{L} \in R^{C\times H\times W}$ with same channels as $f_\mathcal{A}$ are used to encode finer local context by leveraging the spatial local affinity information. Given $f_\mathcal{L}^p$, $f_\mathcal{L}^q\in R^C, p,q\in\{1,...,HW\}$ denote the features in the $p$th and $q$th position of the feature maps $f_\mathcal{L}$. The normalized similarity of local position features $f_\mathcal{L}^p$ and $f_\mathcal{L}^q$ can be obtained by softmax function:
\begin{align}
& s(p,q) = \frac{exp{(d(f_\mathcal{L}^p,f_\mathcal{L}^q))}}{\sum_{q=1}^{HW}exp{(d(f_\mathcal{L}^p,f_\mathcal{L}^q))}}
\end{align}
where $d(\cdot)$ denotes the metric function, in our work, we adopt the inner product of vectors as the metric. By calculating the similarity between any two spatial positions, a similarity matrix $S\in R^{HW\times HW}$ can be obtained. Therefore, the features $f_\mathcal{L}$ can be updated by utilizing the similarity $S$ and $f_\mathcal{A}$ as follows:
\begin{align}
& f_\mathcal{L}^p = \sum_{q=1}^{HW}(s(p,q)\cdot f_\mathcal{L}^q)+f_\mathcal{A}^p
\end{align}
where $f_\mathcal{L}^p$ refers to the output of $p$th position of the CGL module. 
Therefore, the CGL module further aggregates the global and local features, and encodes more contextual relationships over local features.

\subsection{Joint optimization with auxiliary loss}

In the decoder sub-network, we incorporate extracted global context features $f_\mathcal{G}$ and finer local context $f_\mathcal{L}$ to $f_\mathcal{D}$ by concatenation and convolution operation. Then the decoding feature $f_\mathcal{D}$ is followed by a convolution layer and upsampling operator to generate the final prediction map $\tilde{Y}$. The calculation of the prediction mask can be written as:
\begin{eqnarray}
f_\mathcal{D} = \mathcal{F}(\mathcal{C}(f_\mathcal{G},f_\mathcal{L}); \theta_\mathcal{D}) \\
\tilde{Y} = \sigma(Up(\mathcal{F}(f_\mathcal{D}; \theta_{Y})))
\end{eqnarray}
where $\theta_\mathcal{D}$ and $\theta_Y$ are the corresponding convolution parameters. $Up(\cdot)$ denotes the bilinearly interpolation.

As illustrated in Figure \ref{fig:2}, to learn powerful features and facilitate the training process, we add an auxiliary supervised module to the CCA branch and train the whole model by jointly optimizing the losses of two branches in an end-to-end manner. The introduced auxiliary supervised loss helps optimize the learning process while not influencing learning in the main branch. In our work, the weighted binary cross-entropy and dice loss are adopted as our objective function for each branch. For $k$th image, the total objective function is defined as follows:
\begin{align}
& L = \sum_bL_{c}^{(b)} + \lambda L_{d}^{(b)}
\end{align}
where $b$ denotes the different branches (\emph{i.e.} master branch and the auxiliary supervised branch). $\lambda$ represents the balance parameters and is empirically set as $1.0$. $L_c^{(b)}$ and $L_d^{(b)}$ denote the weighted binary cross-entropy (WBCE) loss function and dice loss respectively. The WBCE measures the accuracy of the segmentation, which are defined as follows:
\begin{align}
L_{c} = -\frac{1}{N}[\alpha\sum_{j=1}^{|Y_+|} log P(Y_j^{(k)}=1|X^{(k)};\theta) \nonumber\\
+ (1-\alpha)\sum_{j=1}^{|Y_-|} log P(Y_j^{(k)}=0|X^{(k)};\theta)]
\end{align}
And $L_d$ is the dice loss which is suitable for imbalance samples and measures the coincidence of ground truth annotations and prediction results:
\begin{align}
L_{d} = 1-\frac{2\sum_{j=1}^NY_j^{(k)}\tilde{Y_j}^{(k)} + \epsilon}{\sum_{j=1}^N(Y_j^{(k)})^2 + \sum_{j=1}^N(\tilde{Y_j}^{(k)})^2 + \epsilon}
\end{align}
where $P(Y_j^{(k)}=1|X^{(k)};\theta)$ represents the prediction map in which each value indicates the foreground confidence for $j$th pixel. $\tilde{Y_j}^{(k)}$ is the corresponding prediction value. $|Y_+|$ and $|Y_-|$ mean the number of lesion pixels and non-lesion pixels in ground truth. $\alpha = \frac{|Y_-|}{|Y_+|+|Y_-|}$ refers to the weight. $\theta$ denotes the parameters of our model. $N$ is the total number of pixels. And $\epsilon \in [0,1]$ is a regularization constant for preventing the divide-by-zero risk.

\section{Experiments}

\subsection{Materials}

In our experiments, four skin lesion datasets are employed to validate our algorithm, \emph{i.e} ISIC-2016, ISIC-2017, ISIC-2018, and PH2 datasets.

The first three datasets are collected from the Skin Lesion Analysis Towards Melanoma Detection Challenge, which contains multi-source dermoscopy images of common pigmented skin lesions. For the ISIC-2016 dataset, it includes a training set with $900$ annotated dermoscopy images and a total number of $379$ images for testing. For the ISIC-2017 dataset, it contains $2000$ training images, and a testing set with $600$ images. The publicly available dataset of the ISIC-2018 consists of $2594$ images. In our experiment, we randomly divide $80\%$ images as the training set, and the remaining $20\%$ are as the test set for this dataset.
The PH2 public dataset contains $200$ dermoscopy images. All four datasets are annotated by dermatologists. For the PH2, as the limited of fewer sample numbers, we directly test our model (trained on the ISIC-2017) on the entire PH2 dataset following \citep{bi2019step}.

\subsection{Reference model}

To evaluate the performance of the proposed model, some state-of-the-art medical image segmentation approaches including MultiResUNet \citep{IbtehazRahman2020}, AG-Net \citep{Schlemperetal2019}, CE-Net \citep{Guetal2019}, and related skin lesion segmentation methods are chosen for comparison. Moreover, the top-$5$ algorithms of the ISIC-2016 and ISIC-2017 challenges are also selected. For fair comparisons, the compared results of these methods are generated by the original code released by the authors or taken from their respective publications.

\subsection{Evaluation criteria}

In this paper, five widely accepted metrics are employed as assessment criteria following the ISIC skin lesion segmentation challenges. They include Dice Similarity Coefficient (DI), Jaccard Index (JA), Accuracy (AC), Sensitivity (SE), and Specificity (SP). The details are as follows:
\begin{align}
& DI = \frac{2\cdot TP}{2\cdot TP + FN + FP}, \quad SE = \frac{TP}{TP + FN} \nonumber \\
& JA = \frac{TP}{TP + FN + FP}, \quad \quad SP = \frac{TN} {TN + FP} \nonumber \\
& AC = \frac{TP + TN}{TP + FP + TN + FN}
\end{align}
where $TP$, $TN$, $FP$, and $FN$ refer to the number of true positives, true negatives, false positives, and false negatives, respectively. They are all defined on the pixel level.
Among these metrics, JA mainly assesses the overlapping between predictions and ground truth, which is the most important evaluation metric for segmentation tasks. Also, the participants of challenges are ranked based on the JA. In this work, we mainly refer to JA to evaluate the performance of all approaches.

\subsection{Implementation details}

The proposed method is implemented based on the Pytorch framework and is trained with one NVIDIA TITAN X GPU. In the model, we exploit the standard stochastic gradient descent optimizer with $0.9$ momentum for training. We set the initial learning rate as $10^{-4}$ and adopt the ``poly" learning rate policy following \citep{chen2017rethinking}. The learning rate is multiplied by $(1-\frac{iter}{total_{iter}})^{0.9}$ after each iteration, eventually terminated at $150$ epochs. All the training data is divided into mini-batches for network training, the mini-batch size is set as $8$ during the training stage. In the CCA module, all feature maps are reduced to $256$ channels with convolution operation for each integration. The backbone of the encoder sub-network is based on ResNet with the ASPP module pre-trained on ImageNet. The adopted ASPP consists of one global average pooling, one $1\times 1$ convolution and three $3 \times 3$ convolutions with rates $= (6, 12, 18)$. The output feature map is $1/8$ size of the input image by our model. Then the predicted segmentation masks are bilinearly interpolated to target size directly. All images are uniformly resized into a resolution of $256 \times 256$ for training and testing.

\subsection{Comparisons with the state-of-the-art}

\subsubsection{Evaluation on ISIC-2016 dataset}

\begin{table}[!t]
	\caption{Segmentation performance on ISIC-2016 dataset}
	\label{t1}
	\centering
	\small
	\begin{threeparttable}[b]
		\begin{tabular}{L{4.8cm}C{0.95cm}C{0.95cm}C{0.95cm}C{0.95cm}C{0.95cm}C{0.95cm}}

			\hline
			\hline
			\multirow{2}{*}{\textbf{Methods}} & \multicolumn{5}{c}{\textbf{Averaged evaluation metrics (\%)}} \\
			\cline{2-6}
			& \textbf{\textit{AC}}  & \textbf{\textit{DI}} & \textbf{\textit{JA}} & \textbf{\textit{SE}} & \textbf{\textit{SP}} \\
			\hline
			Team-ExB (*1)         & 95.3    & 91.0   & 84.3   & 91.0   & 96.5  \\
			Team-CUMED (*2)       & 94.9    & 89.7   & 82.9   & 91.1   & 95.7  \\
			Team-test (*3)        & 95.2    & 89.5   & 82.2   & 88.0   & 96.9  \\
			Team-sfu-mial (*4)    & 94.4    & 88.5   & 81.1   & 91.5   & 95.5  \\
			Team-TMU (*5)         & 94.6    & 88.8   & 81.0   & 83.2   & \textbf{98.7}  \\
			FCN \citep{long2015fully}      & 94.1    & 88.6   & 81.3   & 91.7   & 94.9  \\
			\cite{bi2017dermoscopic}      & 95.5    & 91.1   & 84.6   & 92.1   & 96.5  \\
			\cite{yuan2017automatic}      & 95.5    & 91.2   & 84.7   & 91.8   & 96.6  \\
			\cite{yuan2017improving}      & 95.7    & 91.3   & 84.9   & 92.4   & 96.5  \\
			\cite{bi2019step}             & 95.8    & 91.7   & 85.9   & 93.1   & 96.0  \\
			\cite{xie2020skin}            & 93.8    & 91.8   & 85.8   & -      & -     \\
			AG-Net \citep{Schlemperetal2019}         & 95.0    & 89.8   & 83.4   & 92.1   & 95.7  \\
			MultiResUNet \citep{IbtehazRahman2020}   & 95.3    & 91.1   & 84.9   & 92.1   & 96.1  \\
			CE-Net \citep{Guetal2019}                & 95.7    & 91.6   & 85.5   & 92.8   & 95.8  \\
			\hline
			ours      & \textbf{96.1} & \textbf{92.6} & \textbf{87.1}  & \textbf{94.6}  & 96.7 \\
			\hline
			\hline
		\end{tabular}
		\textbf{note:} The *-number indicates the rank of that method in original ISIC-2016 challenge.
	\end{threeparttable}
\end{table}

Table \ref{t1} shows the performance of reference models and our work in the skin lesion segmentation task on the ISIC-2016 dataset. Compared to other segmentation approaches, our model outperforms existing approaches in almost all evaluation metrics. Specifically, the proposed method reaches an overall JA of $87.1\%$ and DI of $92.6\%$. In comparison with the results of the top five teams of the challenge, the proposed architecture achieves JA increases of $2.8\%$ than the best result of the competition ($84.3\%$). Also, we compare the proposed algorithm with other state-of-the-art methods on the testing set. Our approach also obtains improvement among these methods to some extent. Compared with the classical FCN \citep{long2015fully}, our method improves JA by $5.8\%$. And when compared with the recently published works \citep{bi2019step}, \citep{xie2020skin} and \citep{IbtehazRahman2020}, our model has also shown the obvious advantages with $1.2\%$, $1.3\%$, and $2.2\%$ enhancement on metric JA, respectively.

\subsubsection{Evaluation on ISIC-2017 dataset}

\begin{table}[!t]
	\caption{Segmentation performance on ISIC-2017 dataset}
	\label{t2}
	\centering
	\small
	\begin{threeparttable}[b]
		\begin{tabular}{L{4.8cm}C{0.95cm}C{0.95cm}C{0.95cm}C{0.95cm}C{0.95cm}C{0.95cm}}

			\hline
			\hline
			\multirow{2}{*}{\textbf{Methods}} & \multicolumn{5}{c}{\textbf{Averaged evaluation metrics (\%)}} \\
			\cline{2-6}
			& \textbf{\textit{AC}}  & \textbf{\textit{DI}} & \textbf{\textit{JA}} & \textbf{\textit{SE}} & \textbf{\textit{SP}} \\
			\hline
			Team-Mt.Sinai (*1)      & 93.4    & 84.9   & 76.5   & 82.5   & 97.5  \\
			Team-NLP LOGIX (*2)     & 93.2    & 84.7   & 76.2   & 82.0   & 97.8  \\
			Team-BMIT (*3)          & 93.4    & 84.4   & 76.0   & 80.2   & \textbf{98.5}  \\
			Team-BMIT (*4)          & 93.4    & 84.2   & 75.8   & 80.1   & 98.4  \\
			Team-RECOD Titans (*5)  & 93.1    & 83.9   & 75.4   & 81.7   & 97.0  \\
			U-Net \citep{ronneberger2015u}                  & 92.6    & 82.2   & 74.1   & -      & -     \\
			\cite{navarro2018accurate}         & \textbf{95.5}    & 85.4   & 76.9   & -      & -     \\
			\cite{li2018dense}              & 93.9    & 86.6   & 76.5   & 82.5   & 98.4  \\
			\cite{mirikharaji2018deep}     & 93.8    & 85.7   & 77.3   & 85.5   & 96.6  \\
			\cite{bi2019improving}          & -       & 85.1   & 77.1   & -      & -     \\
		  \cite{bi2019step}          & 94.0    & 85.6   & 77.7   & 86.2   & 96.7  \\
			\cite{sarker2018slsdeep}          & 93.6    & 87.8   & 78.2   & 81.6   & 98.3  \\
			\cite{xue2018adversarial}             & 94.1    & 86.7   & 78.5   & -      & -     \\
			\cite{chen2018multi}            & 94.4    & 86.8   & 78.7   & -      & -     \\
			FocusNet \citep{kaul2019focusnet}   & 92.1    & 83.2   & 75.6   & -      & -     \\
			\cite{xie2020skin}            & 93.8    & 86.2   & 78.3   & -      & -     \\
			DSNet \citep{hasan2020dsnet} & -   & -   & 77.5   & 87.5   & 95.5     \\
			DAGAN \citep{lei2020skin}  & 93.5  & 85.9   & 77.1   & 83.5   & 97.6     \\
			AG-Net \citep{Schlemperetal2019}         & 93.5    & 85.3   & 76.9   & 83.5   & 97.4  \\
			MultiResUNet \citep{IbtehazRahman2020}   & 93.6    & 85.2   & 76.8   & 83.9   & 96.8  \\
			CE-Net \citep{Guetal2019}                & 94.0    & 86.5   & 78.5   & 86.9   & 96.4  \\
			\hline
			ours                  & 94.6    & \textbf{87.8}   & \textbf{80.3}  & \textbf{88.6}  & 96.4 \\
			\hline
			\hline
		\end{tabular}
		\textbf{note:} The *-number indicates the rank of that method in original ISIC-2017 challenge.
	\end{threeparttable}
\end{table}

Table \ref{t2} compares the related metrics of reference models and our work in the ISIC-2017 segmentation task. As shown in the table, we can find that the proposed method has the highest DI of $87.8\%$ and the best JA of $80.3\%$ than other methods. In comparison with the results of classical U-Net, our work exceeds it $6.2\%$ on metric JA, and has a $3.8\%$ improvement than the best result of the competition ($76.5\%$). Similarly, our work has considerable margin in terms of both JA and DI compared with some recently published methods, \cite{lei2020skin} (based on GANs) and \cite{chen2018multi} (based on multi-task learning), which obtains $3.2\%$ and $1.6\%$ gains of JA respectively. Overall, the above results suggest that our method can effectively extract the context of images and improve the capacity of feature representation.

\begin{table}[!t]
	\caption{Performance for melanoma and non-melanoma cases on ISIC-2016 dataset}
	\label{t2016}
	\centering
	\small
	\begin{threeparttable}[b]
		\begin{tabular}{L{5.6cm}C{1.1cm}C{1.1cm}C{1.1cm}C{1.1cm}}

			\hline
			\hline
			\multirow{2}{*}{\textbf{Methods}} & \multicolumn{2}{c}{\textbf{Melanoma}} & \multicolumn{2}{c}{\textbf{Non-Melanoma}}\\
			\cline{2-5}
		    & \textbf{\textit{DI}} & \textbf{\textit{JA}} & \textbf{\textit{DI}} & \textbf{\textit{JA}} \\
			\hline
			Team-ExB (*1)         & 90.1    & 82.9   & 91.2   & 84.6     \\
			Team-CUMED (*2)       & 90.0    & 82.9   & 89.7   & 83.0     \\
			Team-test (*3)        & 89.9    & 82.7   & 89.4   & 82.0     \\
			Team-sfu-mial (*4)    & 89.4    & 81.9   & 88.3   & 80.9     \\
			Team-TMU (*5)         & 89.7    & 82.3   & 88.5   & 80.7     \\
			\cite{garcia2019segmentation}  & 88.6  & 80.9   & 78.7   & 86.5    \\
			\cite{bi2019step}  & 91.7  & 85.9   & 91.8   & 85.6       \\
			\hline
			ours                   & \textbf{92.4}   & \textbf{86.6}  & \textbf{92.6}  & \textbf{87.1} \\
			\hline
			\hline
		\end{tabular}
		\textbf{note:} The *-number indicates the rank of that method in original ISIC-2016 challenge.
	\end{threeparttable}
\end{table}

\begin{table}[!t]
	\caption{Performance for melanoma and non-melanoma cases on ISIC-2017 dataset}
	\label{t2017}
	\centering
	\small
	\begin{threeparttable}[b]
		\begin{tabular}{L{5.6cm}C{1.1cm}C{1.1cm}C{1.1cm}C{1.1cm}}

			\hline
			\hline
			\multirow{2}{*}{\textbf{Methods}} & \multicolumn{2}{c}{\textbf{Melanoma}} & \multicolumn{2}{c}{\textbf{Non-Melanoma}}\\
			\cline{2-5}
			& \textbf{\textit{DI}} & \textbf{\textit{JA}} & \textbf{\textit{DI}} & \textbf{\textit{JA}} \\
			\hline
			Team-Mt.Sinai (*1)      & 81.0    & 71.2   & 85.8   & 77.8     \\
			Team-NLP LOGIX (*2)     & 79.1    & 68.8   & 86.1   & 78.0     \\
			Team-BMIT (*3)          & 79.6    & 69.3   & 85.5   & 77.6     \\
			Team-BMIT (*4)          & 79.5    & 69.1   & 85.4   & 77.5     \\
			Team-RECOD Titans (*5)  & 79.1    & 68.8   & 85.1   & 77.0     \\
			\cite{garcia2019segmentation}          & 76.6    & 65.8   & 75.9   & 66.7  \\
			\cite{bi2019improving}          & 82.2    & 72.9   & 85.9   & 78.2  \\
			\cite{bi2019step}          & 81.7    & 72.1   & 86.6   & 79.1  \\
			\hline
			ours                   & \textbf{83.1}   & \textbf{74.3}  & \textbf{88.8}  & \textbf{81.2} \\
			\hline
			\hline
		\end{tabular}
		\textbf{note:} The *-number indicates the rank of that method in original ISIC-2017 challenge.
	\end{threeparttable}
\end{table}

In addition, Table \ref{t2016}-\ref{t2017} show the segmentation performance of different methods on melanoma and non-melanoma cases separately, where our model outperforms other approaches in both cases consistently. Because of the high intra-class variation of the lesion, recognition of melanoma is more difficult than non-melanoma cases. From the tables, we can observe that the proposed method achieves considerable gains in both cases, which suggests the effectiveness of our work on melanoma detection.

\begin{figure*}[!t]
	\centering
	\subfigure[]{\includegraphics[width=1\textwidth]{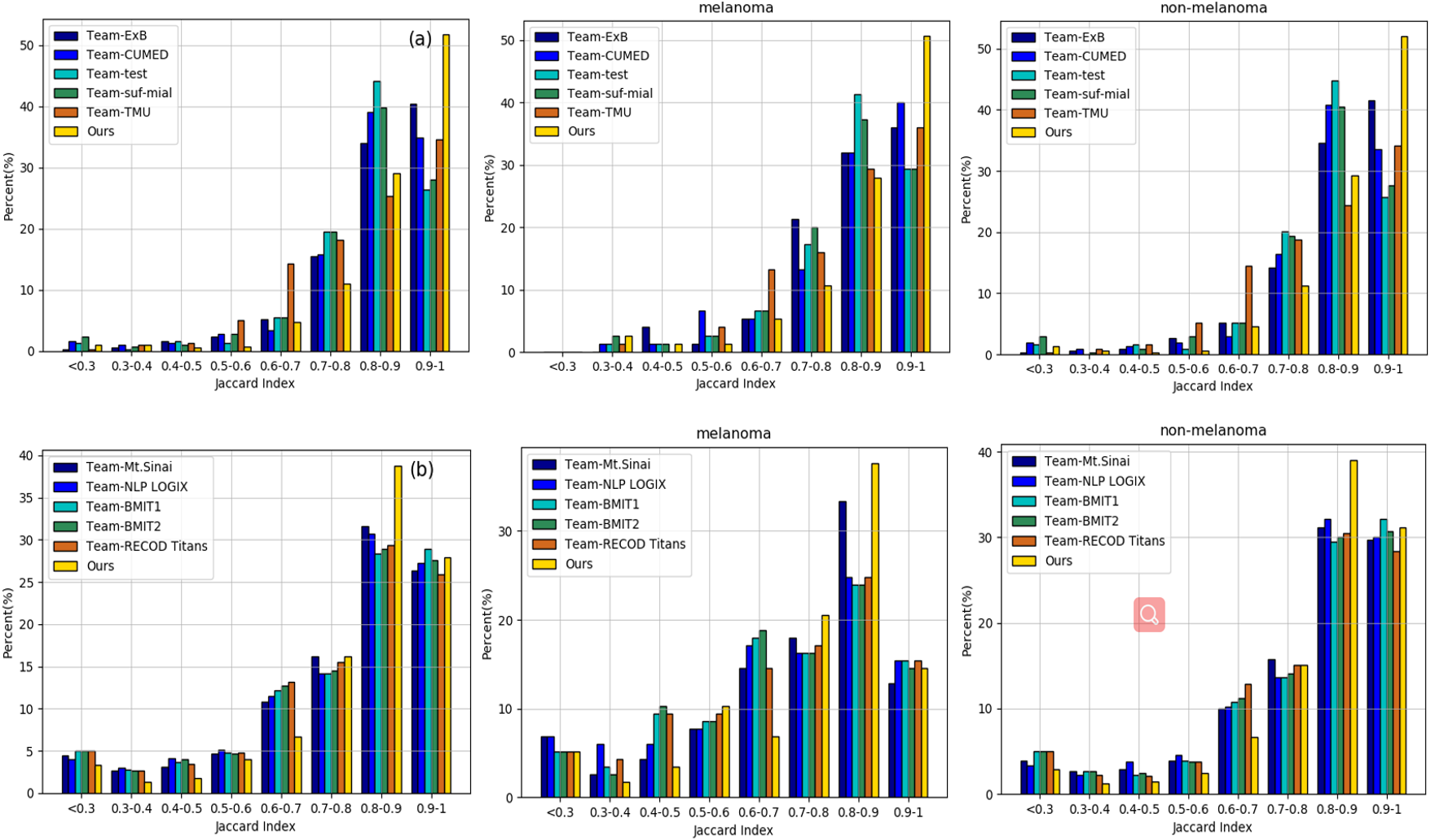}}
	\caption{Distribution of segmentation performance in terms of Jaccard Index for the whole case, melanoma case and non-melanoma case on ISIC-2016 ($1$st row) and ISIC-2017 ($2$nd row) testing dataset.}
	\label{fig:5}
\end{figure*}

Furthermore, Figure \ref{fig:5} provides a clear demonstration of the distribution in terms of JA for the whole case, melanoma case, and non-melanoma case on two datasets. From the figure, we can find that the performances of lesion segmentation with higher JA take up much of these two datasets by our work, which implies better utility and stability than other approaches.

\subsubsection{Evaluation on ISIC-2018 dataset}

\begin{table}[!t]
	\caption{Segmentation performance on ISIC-2018 dataset}
	\label{t2018}
	\centering
	\small
	\begin{threeparttable}[b]
		\begin{tabular}{L{4.8cm}C{0.95cm}C{0.95cm}C{0.95cm}C{0.95cm}C{0.95cm}C{0.95cm}}

			\hline
			\hline
			\multirow{2}{*}{\textbf{Methods}} & \multicolumn{5}{c}{\textbf{Averaged evaluation metrics (\%)}} \\
			\cline{2-6}
			& \textbf{\textit{AC}}  & \textbf{\textit{DI}} & \textbf{\textit{JA}} & \textbf{\textit{SE}} & \textbf{\textit{SP}} \\
			\hline
			FCN \citep{long2015fully}      & 94.5    & 84.9   & 76.6   & -   & -  \\
			U-Net \cite{ronneberger2015u}      & 95.6    & 86.5   & 78.7   & -   & -  \\
			DFN \citep{yu2018learning}    & 94.9    & 86.0   & 77.8   & -      & -     \\
			AG-Net \citep{Schlemperetal2019}         & 95.0    & 87.1   & 79.4   & 86.5   & 97.0  \\
			MultiResUNet \citep{IbtehazRahman2020}   & 95.1    & 88.2   & 80.5   & 86.3   & \textbf{97.1}  \\
			CE-Net \citep{Guetal2019}                & 94.5    & 87.2   & 79.6   & 89.7   & 95.4  \\
			DAGAN \citep{lei2020skin}  & 92.9  & 88.5   & 82.4   & \textbf{95.3}   & 91.1     \\
			CPF-Net \citep{feng2020cpfnet}   & \textbf{96.3}    & 89.8   & 82.8   & -   & - \\
			\hline
			ours      & 95.9 & \textbf{89.9} & \textbf{83.4}  & 93.6  & 96.7 \\
			\hline
			\hline
		\end{tabular}
	\end{threeparttable}
\end{table}

For the ISIC-2018 dataset, our work also achieves state-of-the-art performance with the best average JA score of $83.4\%$. As shown in Table \ref{t2018}, the performances are averagely improved by $4.0\%$ and $3.8\%$ compared with the AG-Net and CE-Net, respectively. Compared to the latest DAGAN and CPF-Net, benefiting from the guidance and integration of context, our method also shows a competitive advantage. This proves that effective long-range contextual information preservation has an important impact on the fine segmentation results.

\subsubsection{Evaluation on PH2 dataset}

\begin{table}[!t]
	\caption{Segmentation performance on PH2 dataset}
	\label{t-ph2}
	\centering
	\small
	\begin{threeparttable}[b]
		\begin{tabular}{L{4.8cm}C{0.95cm}C{0.95cm}C{0.95cm}C{0.95cm}C{0.95cm}C{0.95cm}}

			\hline
			\hline
			\multirow{2}{*}{\textbf{Methods}} & \multicolumn{5}{c}{\textbf{Averaged evaluation metrics (\%)}} \\
			\cline{2-6}
			& \textbf{\textit{AC}}  & \textbf{\textit{DI}} & \textbf{\textit{JA}} & \textbf{\textit{SE}} & \textbf{\textit{SP}} \\
			\hline
			FCN \citep{long2015fully}            & 93.5    & 89.4   & 82.2   & 93.1   & 93.0  \\
			U-Net \cite{ronneberger2015u}          & -       & 87.6   & 78.0   & -      & -     \\
			\cite{bi2017dermoscopic}  & 94.2    & 90.7   & 84.0   & 94.9   & 94.0  \\
			FrCN \citep{al2018skin}           & -       & 91.7   & 84.8   & -      & -     \\
			\citep{bi2019step}   & 95.3    & 92.1   & 85.9   & 96.2   & 94.5  \\
			\citep{xie2020skin}   & 94.9    & 91.9   & 85.7   & -      & -     \\
			\hline
			ours      & \textbf{95.4} & \textbf{92.5} & \textbf{86.6}  & \textbf{98.6}  & \textbf{94.7} \\
			\hline
			\hline
		\end{tabular}
	\end{threeparttable}
\end{table}

Table \ref{t-ph2} shows segmentation results on the PH2 dataset. Our model reaches an overall JA score of $86.6\%$ using the models trained on the ISIC-2017 training set. In comparison with the competitors \cite{bi2019step} and \cite{xie2020skin}, the proposed architecture outnumbers them by $0.7\%$ and $0.9\%$ in JA, respectively. These results prove that our model also has a strong generalization ability.

\subsection{Effect of ablation analysis}

Next, we provide some analysis about our model and give a corresponding comparison on the ISIC-2017 dataset for skin lesion segmentation. Our method employs the proposed CCA module and CGL module assembled in the base encoder-decoder architecture. Therefore, we conduct experiments to validate the usefulness of the proposed method compared with module ablation. In our model, the input features of the CGL module come from generated context features $f_\mathcal{G}$ of the CCA module and the features $f_\mathcal{A}$ of the ASPP module. Without the CCA module, the local contextual information is mined only according to $f_\mathcal{A}$. As shown in Table \ref{t4}, the local affinity module improves the segmentation performance remarkably ($79.5\%$). Compared with the baseline (ResNet+ASPP), only employing the local affinity features in the encoder sub-network yields a $1.1\%$ and $1.6\%$ improvement of DI and JA. Similarly, only adopting the CCA module in our base model, there are $1.1\%$ and $1.5\%$ enhancement on metrics DI and JA than the baseline, respectively. Further, when we integrate the CCA module to extract the richer context and guide the local affinity computation, the proposed method reaches an overall DI of $87.5\%$ and JA of $80.0\%$ without considering the auxiliary loss.

\begin{figure*}[!t]
	\centerline{\includegraphics[width=1\textwidth]{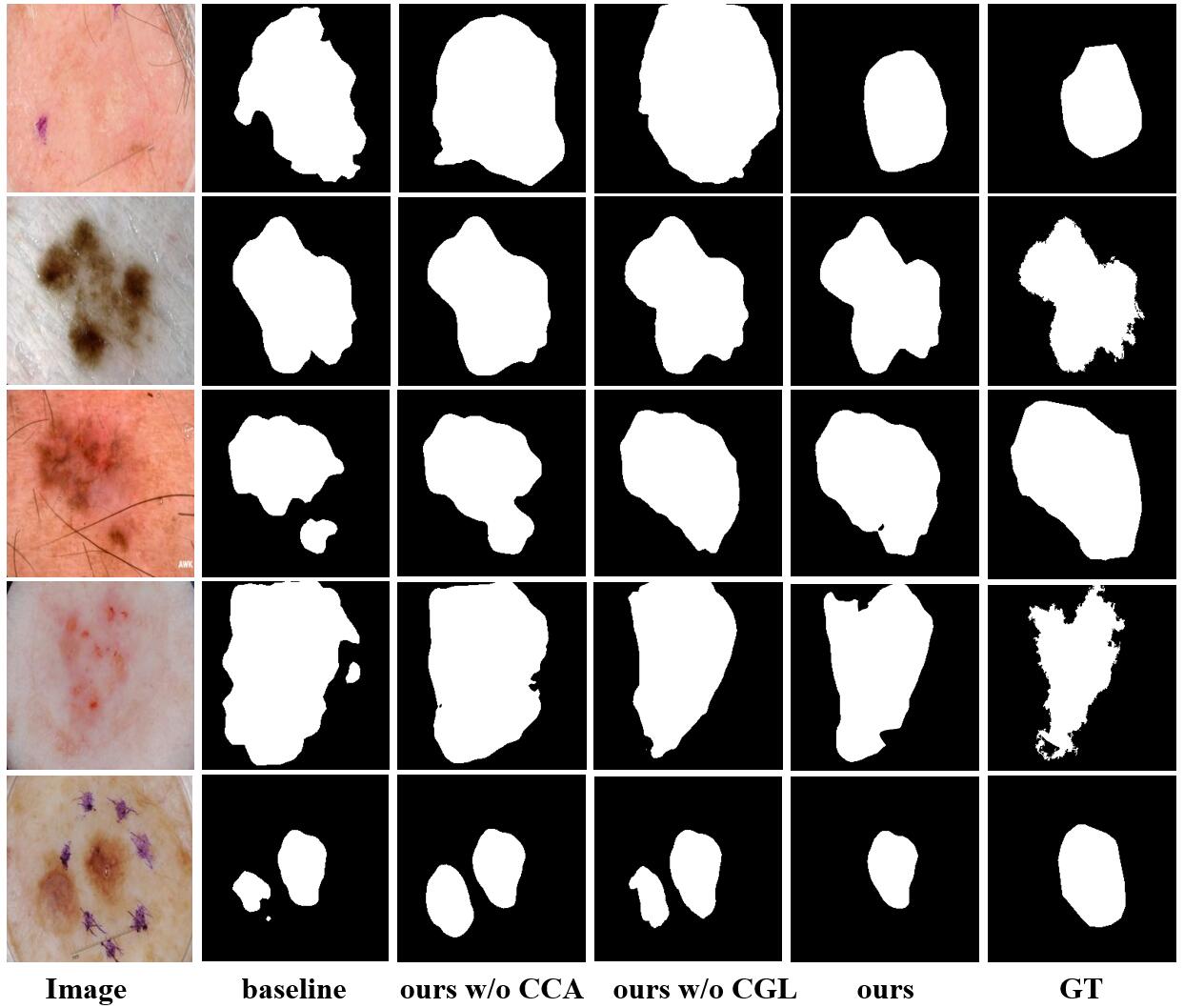}}
	\caption{Visualization of the skin lesion segmentation examples on ISIC-2017 dataset, where the images in the first column are the original dermoscopy images, the segmentation images from the second to the fifth columns are the results using baseline, ours w/o CCA module, ours w/o CGL module and the proposed method respectively, the last column indicates the ground truth.}
	\label{fig:6}
\end{figure*}

Figure \ref{fig:6} also shows the segmentation performances with different modules embedded. From the left to right column, the whole segmentation performances get better with the introduction of local affinity relationship and global context. In particular, the base model can extract coarse region contour on the whole. However, for some images with low contrast or unclear boundary, the segmentation performance is unsatisfactory. Compared with the baseline, the local affinity greatly enhances the expression of features, achieving better local connectivity of the lesion area. But, due to the lack of guidance of global information, many pseudo regions are identified as the lesion. With the CCA, the generated context adaptively integrates the original image and multi-level encoding features, making some lesion areas with low contrast recognized. However, the detail of the local area is not captured finely. Therefore, only integrating the CCA or CGL module, the performance of skin lesion boundary recognition is not accurate. With the generated context and the context-guided local relationship mining, some details and lesion region boundaries are clearer. Results show that the CCA module and CGL module bring great benefit to identify the lesion area, and the proposed method performs much better at details and lesion boundaries.

\begin{table}[htbp]
	\centering
	\small
	\begin{threeparttable}[t]
		\caption{\label{t4} Ablation study on ISIC-2017 dataset.}
		\begin{tabular}{C{1.8cm}C{1.8cm}C{1.8cm}C{1.8cm}C{1.8cm}}
			\hline
			\hline
			\multirow{2}{*}{\textbf{CCA}}   &
			\multirow{2}{*}{\textbf{CGL}}   &
			\multirow{2}{*}{\textbf{AL}}   & \multicolumn{2}{c}{\textbf{ISIC-2017}} \\
			\cline{4-5}
			& & & \textbf{\textit{DI}} & \textbf{\textit{JA}} \\
			\hline
			            &             &         & 86.1  & 77.9 \\
			            & \checkmark  &         & 87.2  & 79.5 \\
			 \checkmark &             &         & 87.2  & 79.4 \\
			 \checkmark & \checkmark  &         & 87.5  & 80.0 \\
			 \checkmark & \checkmark  & \checkmark & \textbf{87.8} & \textbf{80.3} \\
			\hline
			\hline
		\end{tabular}
		\textbf{note:} In the econd row, the 'GCL' represents local similarity calculation when the CCA is not used. 'AL' indicates the auxiliary loss.
	\end{threeparttable}
\end{table}

\subsection{Effect of different numbers of training samples}

\begin{table}[htbp]
	\centering
	\small
	\begin{threeparttable}[t]
		\caption{\label{t6} Evaluating model with different numbers of training samples.}
		\begin{tabular}{C{2.5cm}C{1.5cm}C{0.95cm}C{0.95cm}C{0.95cm}C{0.95cm}C{0.95cm}}
			\hline
			\hline
			\multirow{2}{*}{\textbf{Methods}} & \multirow{2}{*}{\textbf{Numbers}} & \multicolumn{5}{c}{\textbf{ISIC-2017}} \\
			\cline{3-7}
			& & \textbf{\textit{AC}}  & \textbf{\textit{DI}} & \textbf{\textit{JA}} & \textbf{\textit{SE}} & \textbf{\textit{SP}} \\
			\hline
			\multirow{5}{*}{baseline} & 100 & 90.2 & 78.8 & 69.7 & 83.6 & 90.5\\
			& 500   & 91.5 & 81.7 & 73.0 & 89.5 & 91.8\\
			& 1000  & 91.6 & 83.6 & 75.5 & 90.5 & 91.6 \\
			& 1500  & 93.5 & 85.8 & 77.6 & 86.4 & 94.4 \\
			& 2000  & 94.2 & 86.1 & 77.9 & 85.2 & 96.2 \\
			\hline
			\multirow{5}{*}{ours} &100 & 90.6 & 79.6 & 70.6 & 84.4 & 90.6 \\
			& 500   & 92.3 & 83.1 & 74.8 & 88.6 & 92.1 \\
			& 1000  & 92.6 & 84.2 & 76.1 & 89.7 & 93.0 \\
			& 1500  & 94.0 & 86.7 & 79.1 & 88.6 & 94.2 \\
			& 2000  & 94.6 & 87.8 & 80.3 & 88.6 & 96.4 \\
			\hline
			\hline

		\end{tabular}
	\end{threeparttable}
\end{table}

In this section, the effect of different numbers of training samples is analyzed on the ISIC-2017 dataset. We conduct additional experiments by setting the number of training samples to $100$, $500$, $1000$, $1500$, and $2000$. Experiment results are shown in Table \ref{t6}. As we can see, the performances of the baseline and our method are generally improved with the growing numbers of training images in that more skin lesion content is captured. When the numbers increase from $100$ to the whole training set, the overall DI and JA of our model obtain $8.2\%$ and $9.7\%$ gains on the testing set respectively. And compared with the baseline, the proposed method consistently provides superior performances under the different numbers of training samples in DI and JA. In particular, when only half of the whole training samples is used for learning, we achieve an average overall JA of $76.1\%$. And the better segmentation performance is obtained using three-quarters of samples ($79.1\%$), which has outperformed other state-of-the-art methods in Table \ref{t2}.

\subsection{Effect of auxiliary loss}

In our model, we add an auxiliary supervision module to the generated context features of the CCA module for assisting network training. To evaluate the effectiveness of the joint multi-loss optimization, we compare it with the model adopting single-loss in the main decoder branch. As shown in Table \ref{t4}, both the average overall DI and JA increase by $0.3\%$ with the auxiliary loss, which suggests that joint multi-loss optimization can improve the segmentation performance of the proposed method.

\section{Conclusion}

In this paper, we present a cascaded context enhancement neural network for automatic skin lesion segmentation, which produces a richer global context and adaptively integrates local contextual information. Specifically, a cascaded context aggregation module with a gate-based information integration approach is proposed in the encoder sub-network to adaptively capture richer long-range dependencies context. At the same time, a context-guided local affinity module is employed to extract discriminative features by leveraging the local context relationship. Compared with the state-of-the-art deep learning methods for skin lesion segmentation, our proposed approach can effectively achieve multi-level feature integration and generate discriminative feature representation. The experimental results convince the superiority of the proposed method. Based on the outstanding performance of our work, in the future, we will extend our work to support various medical image segmentation tasks including the 2D and 3D image datasets.

%


\bibliography{refs}

\begin{thebibliography}{53}
\expandafter\ifx\csname natexlab\endcsname\relax\def\natexlab#1{#1}\fi
\providecommand{\url}[1]{\texttt{#1}}
\providecommand{\href}[2]{#2}
\providecommand{\path}[1]{#1}
\providecommand{\DOIprefix}{doi:}
\providecommand{\ArXivprefix}{arXiv:}
\providecommand{\URLprefix}{URL: }
\providecommand{\Pubmedprefix}{pmid:}
\providecommand{\doi}[1]{\href{http://dx.doi.org/#1}{\path{#1}}}
\providecommand{\Pubmed}[1]{\href{pmid:#1}{\path{#1}}}
\providecommand{\bibinfo}[2]{#2}
\ifx\xfnm\relax \def\xfnm[#1]{\unskip,\space#1}\fi
\bibitem[{Abbas et~al.(2013)Abbas, Celebi, Serrano, Garcia \&
  Ma}]{abbas2013pattern}
\bibinfo{author}{Abbas, Q.}, \bibinfo{author}{Celebi, M.~E.},
  \bibinfo{author}{Serrano, C.}, \bibinfo{author}{Garcia, I.~F.}, \&
  \bibinfo{author}{Ma, G.} (\bibinfo{year}{2013}).
\newblock \bibinfo{title}{Pattern classification of dermoscopy images: A
  perceptually uniform model}.
\newblock {\it \bibinfo{journal}{Pattern Recognit.}\/},  {\it
  \bibinfo{volume}{46}\/}, \bibinfo{pages}{86--97}.
\bibitem[{Al-Masni et~al.(2018)Al-Masni, Al-Antari, Choi, Han \&
  Kim}]{al2018skin}
\bibinfo{author}{Al-Masni, M.~A.}, \bibinfo{author}{Al-Antari, M.~A.},
  \bibinfo{author}{Choi, M.-T.}, \bibinfo{author}{Han, S.-M.}, \&
  \bibinfo{author}{Kim, T.-S.} (\bibinfo{year}{2018}).
\newblock \bibinfo{title}{Skin lesion segmentation in dermoscopy images via
  deep full resolution convolutional networks}.
\newblock {\it \bibinfo{journal}{Comput. Meth. Programs Biomed.}\/},  {\it
  \bibinfo{volume}{162}\/}, \bibinfo{pages}{221--231}.
\bibitem[{Barata et~al.(2017)Barata, Celebi \& Marques}]{barata2017development}
\bibinfo{author}{Barata, C.}, \bibinfo{author}{Celebi, M.~E.}, \&
  \bibinfo{author}{Marques, J.~S.} (\bibinfo{year}{2017}).
\newblock \bibinfo{title}{Development of a clinically oriented system for
  melanoma diagnosis}.
\newblock {\it \bibinfo{journal}{Pattern Recognit.}\/},  {\it
  \bibinfo{volume}{69}\/}, \bibinfo{pages}{270--285}.
\bibitem[{Bi et~al.(2019{\natexlab{a}})Bi, Feng, Fulham \&
  Kim}]{bi2019improving}
\bibinfo{author}{Bi, L.}, \bibinfo{author}{Feng, D.}, \bibinfo{author}{Fulham,
  M.}, \& \bibinfo{author}{Kim, J.} (\bibinfo{year}{2019}{\natexlab{a}}).
\newblock \bibinfo{title}{Improving skin lesion segmentation via stacked
  adversarial learning}.
\newblock In {\it \bibinfo{booktitle}{Proceedings of the IEEE International
  Symposium on Biomedical Imaging}\/} (pp. \bibinfo{pages}{1100--1103}).
\bibitem[{Bi et~al.(2019{\natexlab{b}})Bi, Kim, Ahn, Kumar, Feng \&
  Fulham}]{bi2019step}
\bibinfo{author}{Bi, L.}, \bibinfo{author}{Kim, J.}, \bibinfo{author}{Ahn, E.},
  \bibinfo{author}{Kumar, A.}, \bibinfo{author}{Feng, D.}, \&
  \bibinfo{author}{Fulham, M.} (\bibinfo{year}{2019}{\natexlab{b}}).
\newblock \bibinfo{title}{Step-wise integration of deep class-specific learning
  for dermoscopic image segmentation}.
\newblock {\it \bibinfo{journal}{Pattern Recognit.}\/},  {\it
  \bibinfo{volume}{85}\/}, \bibinfo{pages}{78--89}.
\bibitem[{Bi et~al.(2017)Bi, Kim, Ahn, Kumar, Fulham \&
  Feng}]{bi2017dermoscopic}
\bibinfo{author}{Bi, L.}, \bibinfo{author}{Kim, J.}, \bibinfo{author}{Ahn, E.},
  \bibinfo{author}{Kumar, A.}, \bibinfo{author}{Fulham, M.}, \&
  \bibinfo{author}{Feng, D.} (\bibinfo{year}{2017}).
\newblock \bibinfo{title}{Dermoscopic image segmentation via multistage fully
  convolutional networks}.
\newblock {\it \bibinfo{journal}{IEEE Trans. Biomed. Eng.}\/},  {\it
  \bibinfo{volume}{64}\/}, \bibinfo{pages}{2065--2074}.
\bibitem[{Cakir et~al.(2012)Cakir, Adamson \& Cingi}]{cakir2012epidemiology}
\bibinfo{author}{Cakir, B.~{\"O}.}, \bibinfo{author}{Adamson, P.}, \&
  \bibinfo{author}{Cingi, C.} (\bibinfo{year}{2012}).
\newblock \bibinfo{title}{Epidemiology and economic burden of nonmelanoma skin
  cancer.}
\newblock {\it \bibinfo{journal}{Facial Plast. Surg. Clin. N. Am.}\/},  {\it
  \bibinfo{volume}{20}\/}, \bibinfo{pages}{419--422}.
\bibitem[{Chen et~al.(2017{\natexlab{a}})Chen, Papandreou, Kokkinos, Murphy \&
  Yuille}]{chen2017deeplab}
\bibinfo{author}{Chen, L.-C.}, \bibinfo{author}{Papandreou, G.},
  \bibinfo{author}{Kokkinos, I.}, \bibinfo{author}{Murphy, K.}, \&
  \bibinfo{author}{Yuille, A.~L.} (\bibinfo{year}{2017}{\natexlab{a}}).
\newblock \bibinfo{title}{Deeplab: Semantic image segmentation with deep
  convolutional nets, atrous convolution, and fully connected crfs}.
\newblock {\it \bibinfo{journal}{IEEE Trans. Pattern Anal. Mach. Intell.}\/},
  {\it \bibinfo{volume}{40}\/}, \bibinfo{pages}{834--848}.
\bibitem[{Chen et~al.(2017{\natexlab{b}})Chen, Papandreou, Schroff \&
  Adam}]{chen2017rethinking}
\bibinfo{author}{Chen, L.-C.}, \bibinfo{author}{Papandreou, G.},
  \bibinfo{author}{Schroff, F.}, \& \bibinfo{author}{Adam, H.}
  (\bibinfo{year}{2017}{\natexlab{b}}).
\newblock \bibinfo{title}{Rethinking atrous convolution for semantic image
  segmentation}.
\newblock {\it \bibinfo{journal}{arXiv preprint arXiv:1706.05587}\/}, .
\bibitem[{Chen et~al.(2018)Chen, Wang, Shi, Liu \& Yu}]{chen2018multi}
\bibinfo{author}{Chen, S.}, \bibinfo{author}{Wang, Z.}, \bibinfo{author}{Shi,
  J.}, \bibinfo{author}{Liu, B.}, \& \bibinfo{author}{Yu, N.}
  (\bibinfo{year}{2018}).
\newblock \bibinfo{title}{A multi-task framework with feature passing module
  for skin lesion classification and segmentation}.
\newblock In {\it \bibinfo{booktitle}{Proceedings of the IEEE International
  Symposium on Biomedical Imaging}\/} (pp. \bibinfo{pages}{1126--1129}).
\bibitem[{Chung \& Sapiro(2000)}]{chung2000segmenting}
\bibinfo{author}{Chung, D.~H.}, \& \bibinfo{author}{Sapiro, G.}
  (\bibinfo{year}{2000}).
\newblock \bibinfo{title}{Segmenting skin lesions with
  partial-differential-equations-based image processing algorithms}.
\newblock {\it \bibinfo{journal}{IEEE Trans. Med. Imaging}\/},  {\it
  \bibinfo{volume}{19}\/}, \bibinfo{pages}{763--767}.
\bibitem[{Codella et~al.(2019)Codella, Rotemberg, Tschandl, Celebi, Dusza,
  Gutman, Helba, Kalloo, Liopyris, Marchetti et~al.}]{codella2019skin}
\bibinfo{author}{Codella, N.}, \bibinfo{author}{Rotemberg, V.},
  \bibinfo{author}{Tschandl, P.}, \bibinfo{author}{Celebi, M.~E.},
  \bibinfo{author}{Dusza, S.}, \bibinfo{author}{Gutman, D.},
  \bibinfo{author}{Helba, B.}, \bibinfo{author}{Kalloo, A.},
  \bibinfo{author}{Liopyris, K.}, \bibinfo{author}{Marchetti, M.} et~al.
  (\bibinfo{year}{2019}).
\newblock \bibinfo{title}{Skin lesion analysis toward melanoma detection 2018:
  A challenge hosted by the international skin imaging collaboration (isic)}.
\newblock {\it \bibinfo{journal}{arXiv preprint arXiv:1902.03368}\/}, .
\bibitem[{Codella et~al.(2018)Codella, Gutman, Celebi, Helba, Marchetti, Dusza,
  Kalloo, Liopyris, Mishra, Kittler et~al.}]{codella2018skin}
\bibinfo{author}{Codella, N.~C.}, \bibinfo{author}{Gutman, D.},
  \bibinfo{author}{Celebi, M.~E.}, \bibinfo{author}{Helba, B.},
  \bibinfo{author}{Marchetti, M.~A.}, \bibinfo{author}{Dusza, S.~W.},
  \bibinfo{author}{Kalloo, A.}, \bibinfo{author}{Liopyris, K.},
  \bibinfo{author}{Mishra, N.}, \bibinfo{author}{Kittler, H.} et~al.
  (\bibinfo{year}{2018}).
\newblock \bibinfo{title}{Skin lesion analysis toward melanoma detection: A
  challenge at the 2017 international symposium on biomedical imaging (isbi),
  hosted by the international skin imaging collaboration (isic)}.
\newblock In {\it \bibinfo{booktitle}{Proceedings of the IEEE International
  Symposium on Biomedical Imaging}\/} (pp. \bibinfo{pages}{168--172}).
\bibitem[{Esteva et~al.(2017)Esteva, Kuprel, Novoa, Ko, Swetter, Blau \&
  Thrun}]{esteva2017dermatologist}
\bibinfo{author}{Esteva, A.}, \bibinfo{author}{Kuprel, B.},
  \bibinfo{author}{Novoa, R.~A.}, \bibinfo{author}{Ko, J.},
  \bibinfo{author}{Swetter, S.~M.}, \bibinfo{author}{Blau, H.~M.}, \&
  \bibinfo{author}{Thrun, S.} (\bibinfo{year}{2017}).
\newblock \bibinfo{title}{Dermatologist-level classification of skin cancer
  with deep neural networks}.
\newblock {\it \bibinfo{journal}{Nature}\/},  {\it \bibinfo{volume}{542}\/},
  \bibinfo{pages}{115--118}.
\bibitem[{Feng et~al.(2020)Feng, Zhao, Shi, Cheng, Wang, Ma, Xiang, Zhu \&
  Chen}]{feng2020cpfnet}
\bibinfo{author}{Feng, S.}, \bibinfo{author}{Zhao, H.}, \bibinfo{author}{Shi,
  F.}, \bibinfo{author}{Cheng, X.}, \bibinfo{author}{Wang, M.},
  \bibinfo{author}{Ma, Y.}, \bibinfo{author}{Xiang, D.}, \bibinfo{author}{Zhu,
  W.}, \& \bibinfo{author}{Chen, X.} (\bibinfo{year}{2020}).
\newblock \bibinfo{title}{Cpfnet: Context pyramid fusion network for medical
  image segmentation}.
\newblock {\it \bibinfo{journal}{IEEE Trans. Med. Imaging}\/},  {\it
  \bibinfo{volume}{39}\/}, \bibinfo{pages}{3008--3018}.
\bibitem[{Ganster et~al.(2001)Ganster, Pinz, Rohrer, Wildling, Binder \&
  Kittler}]{ganster2001automated}
\bibinfo{author}{Ganster, H.}, \bibinfo{author}{Pinz, P.},
  \bibinfo{author}{Rohrer, R.}, \bibinfo{author}{Wildling, E.},
  \bibinfo{author}{Binder, M.}, \& \bibinfo{author}{Kittler, H.}
  (\bibinfo{year}{2001}).
\newblock \bibinfo{title}{Automated melanoma recognition}.
\newblock {\it \bibinfo{journal}{IEEE Trans. Med. Imaging}\/},  {\it
  \bibinfo{volume}{20}\/}, \bibinfo{pages}{233--239}.
\bibitem[{Garcia-Arroyo \& Garcia-Zapirain(2019)}]{garcia2019segmentation}
\bibinfo{author}{Garcia-Arroyo, J.~L.}, \& \bibinfo{author}{Garcia-Zapirain,
  B.} (\bibinfo{year}{2019}).
\newblock \bibinfo{title}{Segmentation of skin lesions in dermoscopy images
  using fuzzy classification of pixels and histogram thresholding}.
\newblock {\it \bibinfo{journal}{Comput. Meth. Programs Biomed.}\/},  {\it
  \bibinfo{volume}{168}\/}, \bibinfo{pages}{11--19}.
\bibitem[{Gu et~al.(2019)Gu, Cheng, Fu, Zhou, Hao, Zhao, Zhang, Gao \&
  Liu}]{Guetal2019}
\bibinfo{author}{Gu, Z.}, \bibinfo{author}{Cheng, J.}, \bibinfo{author}{Fu,
  H.}, \bibinfo{author}{Zhou, K.}, \bibinfo{author}{Hao, H.},
  \bibinfo{author}{Zhao, Y.}, \bibinfo{author}{Zhang, T.},
  \bibinfo{author}{Gao, S.}, \& \bibinfo{author}{Liu, J.}
  (\bibinfo{year}{2019}).
\newblock \bibinfo{title}{Ce-net: context encoder network for 2d medical image
  segmentation}.
\newblock {\it \bibinfo{journal}{IEEE Trans. Med. Imaging}\/},  {\it
  \bibinfo{volume}{38}\/}, \bibinfo{pages}{2281--2292}.
\bibitem[{Gutman et~al.(2016)Gutman, Codella, Celebi, Helba, Marchetti, Mishra
  \& Halpern}]{gutman2016skin}
\bibinfo{author}{Gutman, D.}, \bibinfo{author}{Codella, N.~C.},
  \bibinfo{author}{Celebi, E.}, \bibinfo{author}{Helba, B.},
  \bibinfo{author}{Marchetti, M.}, \bibinfo{author}{Mishra, N.}, \&
  \bibinfo{author}{Halpern, A.} (\bibinfo{year}{2016}).
\newblock \bibinfo{title}{Skin lesion analysis toward melanoma detection: A
  challenge at the international symposium on biomedical imaging (isbi) 2016,
  hosted by the international skin imaging collaboration (isic)}.
\newblock {\it \bibinfo{journal}{arXiv preprint arXiv:1605.01397}\/}, .
\bibitem[{Guy~Jr et~al.(2015)Guy~Jr, Machlin, Ekwueme \&
  Yabroff}]{guy2015prevalence}
\bibinfo{author}{Guy~Jr, G.~P.}, \bibinfo{author}{Machlin, S.~R.},
  \bibinfo{author}{Ekwueme, D.~U.}, \& \bibinfo{author}{Yabroff, K.~R.}
  (\bibinfo{year}{2015}).
\newblock \bibinfo{title}{Prevalence and costs of skin cancer treatment in the
  us, 2002- 2006 and 2007- 2011}.
\newblock {\it \bibinfo{journal}{Am. J. Prev. Med.}\/},  {\it
  \bibinfo{volume}{48}\/}, \bibinfo{pages}{183--187}.
\bibitem[{Hasan et~al.(2020)Hasan, Dahal, Samarakoon, Tushar \&
  Mart{\'\i}}]{hasan2020dsnet}
\bibinfo{author}{Hasan, M.~K.}, \bibinfo{author}{Dahal, L.},
  \bibinfo{author}{Samarakoon, P.~N.}, \bibinfo{author}{Tushar, F.~I.}, \&
  \bibinfo{author}{Mart{\'\i}, R.} (\bibinfo{year}{2020}).
\newblock \bibinfo{title}{Dsnet: Automatic dermoscopic skin lesion
  segmentation}.
\newblock {\it \bibinfo{journal}{Comput. Biol. Med.}\/},  {\it
  \bibinfo{volume}{120}\/}, \bibinfo{pages}{103738}.
\bibitem[{He et~al.(2016)He, Zhang, Ren \& Sun}]{he2016deep}
\bibinfo{author}{He, K.}, \bibinfo{author}{Zhang, X.}, \bibinfo{author}{Ren,
  S.}, \& \bibinfo{author}{Sun, J.} (\bibinfo{year}{2016}).
\newblock \bibinfo{title}{Deep residual learning for image recognition}.
\newblock In {\it \bibinfo{booktitle}{Proceedings of the IEEE Conference on
  Computer Vision and Pattern Recognition}\/} (pp. \bibinfo{pages}{770--778}).
\bibitem[{Ibtehaz \& Rahman(2020)}]{IbtehazRahman2020}
\bibinfo{author}{Ibtehaz, N.}, \& \bibinfo{author}{Rahman, M.}
  (\bibinfo{year}{2020}).
\newblock \bibinfo{title}{Multiresunet: Rethinking the u-net architecture for
  multimodal biomedical image segmentation}.
\newblock {\it \bibinfo{journal}{Neural Networks}\/},  {\it
  \bibinfo{volume}{121}\/}, \bibinfo{pages}{74--87}.
\bibitem[{Kaul et~al.(2019)Kaul, Manandhar \& Pears}]{kaul2019focusnet}
\bibinfo{author}{Kaul, C.}, \bibinfo{author}{Manandhar, S.}, \&
  \bibinfo{author}{Pears, N.} (\bibinfo{year}{2019}).
\newblock \bibinfo{title}{Focusnet: An attention-based fully convolutional
  network for medical image segmentation}.
\newblock In {\it \bibinfo{booktitle}{Proceedings of the IEEE International
  Symposium on Biomedical Imaging}\/} (pp. \bibinfo{pages}{455--458}).
\bibitem[{Kittler et~al.(2002)Kittler, Pehamberger, Wolff \&
  Binder}]{kittler2002diagnostic}
\bibinfo{author}{Kittler, H.}, \bibinfo{author}{Pehamberger, H.},
  \bibinfo{author}{Wolff, K.}, \& \bibinfo{author}{Binder, M.}
  (\bibinfo{year}{2002}).
\newblock \bibinfo{title}{Diagnostic accuracy of dermoscopy}.
\newblock {\it \bibinfo{journal}{The lancet oncology}\/},  {\it
  \bibinfo{volume}{3}\/}, \bibinfo{pages}{159--165}.
\bibitem[{Leachman et~al.(2016)Leachman, Cassidy, Chen, Curiel, Geller, Gareau,
  Pellacani, Grichnik, Malvehy, North et~al.}]{leachman2016methods}
\bibinfo{author}{Leachman, S.~A.}, \bibinfo{author}{Cassidy, P.~B.},
  \bibinfo{author}{Chen, S.~C.}, \bibinfo{author}{Curiel, C.},
  \bibinfo{author}{Geller, A.}, \bibinfo{author}{Gareau, D.},
  \bibinfo{author}{Pellacani, G.}, \bibinfo{author}{Grichnik, J.~M.},
  \bibinfo{author}{Malvehy, J.}, \bibinfo{author}{North, J.} et~al.
  (\bibinfo{year}{2016}).
\newblock \bibinfo{title}{Methods of melanoma detection}.
\newblock {\it \bibinfo{journal}{Melanoma}\/},  (pp. \bibinfo{pages}{51--105}).
\bibitem[{Lei et~al.(2020)Lei, Xia, Jiang, Jiang, Ge, Xu, Qin, Chen, Wang \&
  Wang}]{lei2020skin}
\bibinfo{author}{Lei, B.}, \bibinfo{author}{Xia, Z.}, \bibinfo{author}{Jiang,
  F.}, \bibinfo{author}{Jiang, X.}, \bibinfo{author}{Ge, Z.},
  \bibinfo{author}{Xu, Y.}, \bibinfo{author}{Qin, J.}, \bibinfo{author}{Chen,
  S.}, \bibinfo{author}{Wang, T.}, \& \bibinfo{author}{Wang, S.}
  (\bibinfo{year}{2020}).
\newblock \bibinfo{title}{Skin lesion segmentation via generative adversarial
  networks with dual discriminators}.
\newblock {\it \bibinfo{journal}{Med. Image Anal.}\/},  {\it
  \bibinfo{volume}{64}\/}, \bibinfo{pages}{101716}.
\bibitem[{Lessmann et~al.(2017)Lessmann, van Ginneken, Zreik, de~Jong, de~Vos,
  Viergever \& I{\v{s}}gum}]{lessmann2017automatic}
\bibinfo{author}{Lessmann, N.}, \bibinfo{author}{van Ginneken, B.},
  \bibinfo{author}{Zreik, M.}, \bibinfo{author}{de~Jong, P.~A.},
  \bibinfo{author}{de~Vos, B.~D.}, \bibinfo{author}{Viergever, M.~A.}, \&
  \bibinfo{author}{I{\v{s}}gum, I.} (\bibinfo{year}{2017}).
\newblock \bibinfo{title}{Automatic calcium scoring in low-dose chest ct using
  deep neural networks with dilated convolutions}.
\newblock {\it \bibinfo{journal}{IEEE Trans. Med. Imaging}\/},  {\it
  \bibinfo{volume}{37}\/}, \bibinfo{pages}{615--625}.
\bibitem[{Li et~al.(2018)Li, He, Zhou, Yu, Ni, Chen, Wang \& Lei}]{li2018dense}
\bibinfo{author}{Li, H.}, \bibinfo{author}{He, X.}, \bibinfo{author}{Zhou, F.},
  \bibinfo{author}{Yu, Z.}, \bibinfo{author}{Ni, D.}, \bibinfo{author}{Chen,
  S.}, \bibinfo{author}{Wang, T.}, \& \bibinfo{author}{Lei, B.}
  (\bibinfo{year}{2018}).
\newblock \bibinfo{title}{Dense deconvolutional network for skin lesion
  segmentation}.
\newblock {\it \bibinfo{journal}{IEEE J. Biomed. Health Inform.}\/},  {\it
  \bibinfo{volume}{23}\/}, \bibinfo{pages}{527--537}.
\bibitem[{Lin et~al.(2018)Lin, Ji, Lischinski, Cohen-Or \&
  Huang}]{lin2018multi}
\bibinfo{author}{Lin, D.}, \bibinfo{author}{Ji, Y.},
  \bibinfo{author}{Lischinski, D.}, \bibinfo{author}{Cohen-Or, D.}, \&
  \bibinfo{author}{Huang, H.} (\bibinfo{year}{2018}).
\newblock \bibinfo{title}{Multi-scale context intertwining for semantic
  segmentation}.
\newblock In {\it \bibinfo{booktitle}{Proceedings of the European Conference on
  Computer Vision}\/} (pp. \bibinfo{pages}{603--619}).
\bibitem[{Long et~al.(2015)Long, Shelhamer \& Darrell}]{long2015fully}
\bibinfo{author}{Long, J.}, \bibinfo{author}{Shelhamer, E.}, \&
  \bibinfo{author}{Darrell, T.} (\bibinfo{year}{2015}).
\newblock \bibinfo{title}{Fully convolutional networks for semantic
  segmentation}.
\newblock In {\it \bibinfo{booktitle}{Proceedings of the IEEE Conference on
  Computer Vision and Pattern Recognition}\/} (pp.
  \bibinfo{pages}{3431--3440}).
\bibitem[{Luo et~al.(2017)Luo, Mishra, Achkar, Eichel, Li \&
  Jodoin}]{luo2017non}
\bibinfo{author}{Luo, Z.}, \bibinfo{author}{Mishra, A.},
  \bibinfo{author}{Achkar, A.}, \bibinfo{author}{Eichel, J.},
  \bibinfo{author}{Li, S.}, \& \bibinfo{author}{Jodoin, P.-M.}
  (\bibinfo{year}{2017}).
\newblock \bibinfo{title}{Non-local deep features for salient object
  detection}.
\newblock In {\it \bibinfo{booktitle}{Proceedings of the IEEE Conference on
  Computer Vision and Pattern Recognition}\/} (pp.
  \bibinfo{pages}{6609--6617}).
\bibitem[{Mendon{\c{c}}a et~al.(2013)Mendon{\c{c}}a, Ferreira, Marques, Marcal
  \& Rozeira}]{mendoncca2013ph}
\bibinfo{author}{Mendon{\c{c}}a, T.}, \bibinfo{author}{Ferreira, P.~M.},
  \bibinfo{author}{Marques, J.~S.}, \bibinfo{author}{Marcal, A.~R.}, \&
  \bibinfo{author}{Rozeira, J.} (\bibinfo{year}{2013}).
\newblock \bibinfo{title}{Ph 2-a dermoscopic image database for research and
  benchmarking}.
\newblock In {\it \bibinfo{booktitle}{Proceedings of the Annual International
  Conference of the IEEE Engineering in Medicine and Biology Society}\/} (pp.
  \bibinfo{pages}{5437--5440}).
\bibitem[{Mirikharaji et~al.(2018)Mirikharaji, Izadi, Kawahara \&
  Hamarneh}]{mirikharaji2018deep}
\bibinfo{author}{Mirikharaji, Z.}, \bibinfo{author}{Izadi, S.},
  \bibinfo{author}{Kawahara, J.}, \& \bibinfo{author}{Hamarneh, G.}
  (\bibinfo{year}{2018}).
\newblock \bibinfo{title}{Deep auto-context fully convolutional neural network
  for skin lesion segmentation}.
\newblock In {\it \bibinfo{booktitle}{Proceedings of the IEEE International
  Symposium on Biomedical Imaging}\/} (pp. \bibinfo{pages}{877--880}).
\bibitem[{Navarro et~al.(2018)Navarro, Escudero-Vi{\~n}olo \&
  Besc{\'o}s}]{navarro2018accurate}
\bibinfo{author}{Navarro, F.}, \bibinfo{author}{Escudero-Vi{\~n}olo, M.}, \&
  \bibinfo{author}{Besc{\'o}s, J.} (\bibinfo{year}{2018}).
\newblock \bibinfo{title}{Accurate segmentation and registration of skin lesion
  images to evaluate lesion change}.
\newblock {\it \bibinfo{journal}{IEEE J. Biomed. Health Inform.}\/},  {\it
  \bibinfo{volume}{23}\/}, \bibinfo{pages}{501--508}.
\bibitem[{Ronneberger et~al.(2015)Ronneberger, Fischer \&
  Brox}]{ronneberger2015u}
\bibinfo{author}{Ronneberger, O.}, \bibinfo{author}{Fischer, P.}, \&
  \bibinfo{author}{Brox, T.} (\bibinfo{year}{2015}).
\newblock \bibinfo{title}{U-net: Convolutional networks for biomedical image
  segmentation}.
\newblock In {\it \bibinfo{booktitle}{International Conference on Medical image
  computing and computer-assisted intervention}\/} (pp.
  \bibinfo{pages}{234--241}).
\bibitem[{Russakovsky et~al.(2015)Russakovsky, Deng, Su, Krause, Satheesh, Ma,
  Huang, Karpathy, Khosla, Bernstein et~al.}]{russakovsky2015imagenet}
\bibinfo{author}{Russakovsky, O.}, \bibinfo{author}{Deng, J.},
  \bibinfo{author}{Su, H.}, \bibinfo{author}{Krause, J.},
  \bibinfo{author}{Satheesh, S.}, \bibinfo{author}{Ma, S.},
  \bibinfo{author}{Huang, Z.}, \bibinfo{author}{Karpathy, A.},
  \bibinfo{author}{Khosla, A.}, \bibinfo{author}{Bernstein, M.} et~al.
  (\bibinfo{year}{2015}).
\newblock \bibinfo{title}{Imagenet large scale visual recognition challenge}.
\newblock {\it \bibinfo{journal}{Int. J. Comput. Vis.}\/},  {\it
  \bibinfo{volume}{115}\/}, \bibinfo{pages}{211--252}.
\bibitem[{Sarker et~al.(2018)Sarker, Rashwan, Akram, Banu, Saleh, Singh,
  Chowdhury, Abdulwahab, Romani, Radeva et~al.}]{sarker2018slsdeep}
\bibinfo{author}{Sarker, M. M.~K.}, \bibinfo{author}{Rashwan, H.~A.},
  \bibinfo{author}{Akram, F.}, \bibinfo{author}{Banu, S.~F.},
  \bibinfo{author}{Saleh, A.}, \bibinfo{author}{Singh, V.~K.},
  \bibinfo{author}{Chowdhury, F.~U.}, \bibinfo{author}{Abdulwahab, S.},
  \bibinfo{author}{Romani, S.}, \bibinfo{author}{Radeva, P.} et~al.
  (\bibinfo{year}{2018}).
\newblock \bibinfo{title}{Slsdeep: Skin lesion segmentation based on dilated
  residual and pyramid pooling networks}.
\newblock In {\it \bibinfo{booktitle}{Proceedings of the International
  Conference on Medical Image Computing and Computer-Assisted Intervention}\/}
  (pp. \bibinfo{pages}{21--29}).
\bibitem[{Schlemper et~al.(2019)Schlemper, Oktay, Schaap, Heinrich, Kainz,
  Glocker \& Rueckert}]{Schlemperetal2019}
\bibinfo{author}{Schlemper, J.}, \bibinfo{author}{Oktay, O.},
  \bibinfo{author}{Schaap, M.}, \bibinfo{author}{Heinrich, M.},
  \bibinfo{author}{Kainz, B.}, \bibinfo{author}{Glocker, B.}, \&
  \bibinfo{author}{Rueckert, D.} (\bibinfo{year}{2019}).
\newblock \bibinfo{title}{Attention gated networks: Learning to leverage
  salient regions in medical images}.
\newblock {\it \bibinfo{journal}{Med. Image Anal.}\/},  {\it
  \bibinfo{volume}{53}\/}, \bibinfo{pages}{197--207}.
\bibitem[{Siegel et~al.(2016)Siegel, Miller \& Jemal}]{siegel2016cancer}
\bibinfo{author}{Siegel, R.~L.}, \bibinfo{author}{Miller, K.~D.}, \&
  \bibinfo{author}{Jemal, A.} (\bibinfo{year}{2016}).
\newblock \bibinfo{title}{Cancer statistics, 2016}.
\newblock {\it \bibinfo{journal}{CA-Cancer J. Clin.}\/},  {\it
  \bibinfo{volume}{66}\/}, \bibinfo{pages}{7--30}.
\bibitem[{Silveira et~al.(2009)Silveira, Nascimento, Marques, Mar{\c{c}}al,
  Mendon{\c{c}}a, Yamauchi, Maeda \& Rozeira}]{silveira2009comparison}
\bibinfo{author}{Silveira, M.}, \bibinfo{author}{Nascimento, J.~C.},
  \bibinfo{author}{Marques, J.~S.}, \bibinfo{author}{Mar{\c{c}}al, A.~R.},
  \bibinfo{author}{Mendon{\c{c}}a, T.}, \bibinfo{author}{Yamauchi, S.},
  \bibinfo{author}{Maeda, J.}, \& \bibinfo{author}{Rozeira, J.}
  (\bibinfo{year}{2009}).
\newblock \bibinfo{title}{Comparison of segmentation methods for melanoma
  diagnosis in dermoscopy images}.
\newblock {\it \bibinfo{journal}{IEEE J. Sel. Top. Signal Process.}\/},  {\it
  \bibinfo{volume}{3}\/}, \bibinfo{pages}{35--45}.
\bibitem[{Tang(2009)}]{tang2009multi}
\bibinfo{author}{Tang, J.} (\bibinfo{year}{2009}).
\newblock \bibinfo{title}{A multi-direction gvf snake for the segmentation of
  skin cancer images}.
\newblock {\it \bibinfo{journal}{Pattern Recognit.}\/},  {\it
  \bibinfo{volume}{42}\/}, \bibinfo{pages}{1172--1179}.
\bibitem[{Xie \& Bovik(2013)}]{xie2013automatic}
\bibinfo{author}{Xie, F.}, \& \bibinfo{author}{Bovik, A.~C.}
  (\bibinfo{year}{2013}).
\newblock \bibinfo{title}{Automatic segmentation of dermoscopy images using
  self-generating neural networks seeded by genetic algorithm}.
\newblock {\it \bibinfo{journal}{Pattern Recognit.}\/},  {\it
  \bibinfo{volume}{46}\/}, \bibinfo{pages}{1012--1019}.
\bibitem[{Xie et~al.(2020)Xie, Yang, Liu, Jiang, Zheng \& Wang}]{xie2020skin}
\bibinfo{author}{Xie, F.}, \bibinfo{author}{Yang, J.}, \bibinfo{author}{Liu,
  J.}, \bibinfo{author}{Jiang, Z.}, \bibinfo{author}{Zheng, Y.}, \&
  \bibinfo{author}{Wang, Y.} (\bibinfo{year}{2020}).
\newblock \bibinfo{title}{Skin lesion segmentation using high-resolution
  convolutional neural network}.
\newblock {\it \bibinfo{journal}{Comput. Meth. Programs Biomed.}\/},  {\it
  \bibinfo{volume}{186}\/}, \bibinfo{pages}{105241}.
\bibitem[{Xue et~al.(2018)Xue, Xu \& Huang}]{xue2018adversarial}
\bibinfo{author}{Xue, Y.}, \bibinfo{author}{Xu, T.}, \& \bibinfo{author}{Huang,
  X.} (\bibinfo{year}{2018}).
\newblock \bibinfo{title}{Adversarial learning with multi-scale loss for skin
  lesion segmentation}.
\newblock In {\it \bibinfo{booktitle}{Proceedings of the IEEE International
  Symposium on Biomedical Imaging}\/} (pp. \bibinfo{pages}{859--863}).
\bibitem[{Yang et~al.(2018)Yang, Yu, Zhang, Li \& Yang}]{yang2018denseaspp}
\bibinfo{author}{Yang, M.}, \bibinfo{author}{Yu, K.}, \bibinfo{author}{Zhang,
  C.}, \bibinfo{author}{Li, Z.}, \& \bibinfo{author}{Yang, K.}
  (\bibinfo{year}{2018}).
\newblock \bibinfo{title}{Denseaspp for semantic segmentation in street
  scenes}.
\newblock In {\it \bibinfo{booktitle}{Proceedings of the IEEE Conference on
  Computer Vision and Pattern Recognition}\/} (pp.
  \bibinfo{pages}{3684--3692}).
\bibitem[{Yu et~al.(2018)Yu, Wang, Peng, Gao, Yu \& Sang}]{yu2018learning}
\bibinfo{author}{Yu, C.}, \bibinfo{author}{Wang, J.}, \bibinfo{author}{Peng,
  C.}, \bibinfo{author}{Gao, C.}, \bibinfo{author}{Yu, G.}, \&
  \bibinfo{author}{Sang, N.} (\bibinfo{year}{2018}).
\newblock \bibinfo{title}{Learning a discriminative feature network for
  semantic segmentation}.
\newblock In {\it \bibinfo{booktitle}{Proceedings of the IEEE Conference on
  Computer Vision and Pattern Recognition}\/} (pp.
  \bibinfo{pages}{1857--1866}).
\bibitem[{Yuan et~al.(2009)Yuan, Situ \& Zouridakis}]{yuan2009narrow}
\bibinfo{author}{Yuan, X.}, \bibinfo{author}{Situ, N.}, \&
  \bibinfo{author}{Zouridakis, G.} (\bibinfo{year}{2009}).
\newblock \bibinfo{title}{A narrow band graph partitioning method for skin
  lesion segmentation}.
\newblock {\it \bibinfo{journal}{Pattern Recognit.}\/},  {\it
  \bibinfo{volume}{42}\/}, \bibinfo{pages}{1017--1028}.
\bibitem[{Yuan et~al.(2017)Yuan, Chao \& Lo}]{yuan2017automatic}
\bibinfo{author}{Yuan, Y.}, \bibinfo{author}{Chao, M.}, \& \bibinfo{author}{Lo,
  Y.-C.} (\bibinfo{year}{2017}).
\newblock \bibinfo{title}{Automatic skin lesion segmentation using deep fully
  convolutional networks with jaccard distance}.
\newblock {\it \bibinfo{journal}{IEEE Trans. Med. Imaging}\/},  {\it
  \bibinfo{volume}{36}\/}, \bibinfo{pages}{1876--1886}.
\bibitem[{Yuan \& Lo(2017)}]{yuan2017improving}
\bibinfo{author}{Yuan, Y.}, \& \bibinfo{author}{Lo, Y.-C.}
  (\bibinfo{year}{2017}).
\newblock \bibinfo{title}{Improving dermoscopic image segmentation with
  enhanced convolutional-deconvolutional networks}.
\newblock {\it \bibinfo{journal}{IEEE J. Biomed. Health Inform.}\/},  {\it
  \bibinfo{volume}{23}\/}, \bibinfo{pages}{519--526}.
\bibitem[{Zhang et~al.(2018)Zhang, Zhang, Peng, Xue \& Sun}]{zhang2018exfuse}
\bibinfo{author}{Zhang, Z.}, \bibinfo{author}{Zhang, X.},
  \bibinfo{author}{Peng, C.}, \bibinfo{author}{Xue, X.}, \&
  \bibinfo{author}{Sun, J.} (\bibinfo{year}{2018}).
\newblock \bibinfo{title}{Exfuse: Enhancing feature fusion for semantic
  segmentation}.
\newblock In {\it \bibinfo{booktitle}{Proceedings of the European Conference on
  Computer Vision}\/} (pp. \bibinfo{pages}{269--284}).
\bibitem[{Zhao et~al.(2017)Zhao, Shi, Qi, Wang \& Jia}]{zhao2017pyramid}
\bibinfo{author}{Zhao, H.}, \bibinfo{author}{Shi, J.}, \bibinfo{author}{Qi,
  X.}, \bibinfo{author}{Wang, X.}, \& \bibinfo{author}{Jia, J.}
  (\bibinfo{year}{2017}).
\newblock \bibinfo{title}{Pyramid scene parsing network}.
\newblock In {\it \bibinfo{booktitle}{Proceedings of the IEEE Conference on
  Computer Vision and Pattern Recognition}\/} (pp.
  \bibinfo{pages}{2881--2890}).
\bibitem[{Zhao et~al.(2018)Zhao, Zhang, Liu, Shi, Loy, Lin \&
  Jia}]{zhao2018psanet}
\bibinfo{author}{Zhao, H.}, \bibinfo{author}{Zhang, Y.}, \bibinfo{author}{Liu,
  S.}, \bibinfo{author}{Shi, J.}, \bibinfo{author}{Loy, C.~C.},
  \bibinfo{author}{Lin, D.}, \& \bibinfo{author}{Jia, J.}
  (\bibinfo{year}{2018}).
\newblock \bibinfo{title}{Psanet: Point-wise spatial attention network for
  scene parsing}.
\newblock In {\it \bibinfo{booktitle}{Proceedings of the European Conference on
  Computer Vision}\/} (pp. \bibinfo{pages}{267--283}).

\end{thebibliography}

\end{document}